\documentclass[12pt]{article}
\usepackage{amsmath,amssymb,url,epsfig,color,cite}

\definecolor{verdes}{cmyk}{0.92,0,0.59,0.4}
\ifx\pdfoutput\undefined
\usepackage[dvips,bookmarks=false]{hyperref}	
\else
\usepackage{hyperref}	
\fi
\hypersetup{colorlinks,bookmarksopen,bookmarksnumbered,citecolor=verdes,
linkcolor=blus,pdfstartview=FitH,urlcolor=rossos}

\def\tm{{\tilde m}}
\def\m@th{\mathsurround=0pt }
\def\leftrightarrowfill{$\m@th \mathord\leftarrow \mkern-6mu
        \cleaders\hbox{$\mkern-2mu \mathord- \mkern-2mu$}\hfill
        \mkern-6mu \mathord\rightarrow$}

\def\overleftrightarrow#1{\vbox{\ialign{##\crcr
        \leftrightarrowfill\crcr\noalign{\kern-1pt\nointerlineskip}
        $\hfil\displaystyle{#1}\hfil$\crcr}}}

\definecolor{rosso}{cmyk}{0,1,1,0.4}
\definecolor{rossos}{cmyk}{0,1,1,0.55}
\definecolor{rossoc}{cmyk}{0,1,1,0.2}
\definecolor{blu}{cmyk}{1,1,0,0.3}
\definecolor{blus}{cmyk}{1,1,0,0.6}
\definecolor{bluc}{cmyk}{1,1,0,0.1}
\definecolor{verde}{cmyk}{0.92,0,0.59,0.25}
\definecolor{verdec}{cmyk}{0.92,0,0.59,0.15}
\definecolor{verdes}{cmyk}{0.92,0,0.59,0.4}
\definecolor{grigio}{cmyk}{0,0,0,0.07}
\definecolor{rosa}{cmyk}{0,0.1,0.1,0.02}
\definecolor{rosino}{cmyk}{0,0.05,0.05,0.02}
\definecolor{rosas}{cmyk}{0,0.3,0.25,0.05}
\definecolor{celeste}{cmyk}{0.1,0,0,0.02}
\definecolor{giallino}{cmyk}{0,0,0.4,0.02}
\definecolor{rosso}{cmyk}{0,1,1,0.4}
\definecolor{rossos}{cmyk}{0,1,1,0.55}
\definecolor{rossoc}{cmyk}{0,1,1,0.2}
\definecolor{blu}{cmyk}{1,1,0,0.3}
\definecolor{bluc}{cmyk}{1,1,0,0.1}
\definecolor{blucc}{cmyk}{0.7,0.5,0,0}
\definecolor{viola}{cmyk}{0,1,0,0.6}
\definecolor{viola2}{cmyk}{0,1,0.2,0.6}
\definecolor{verde}{cmyk}{0.92,0,0.59,0.25}
\definecolor{verdec}{cmyk}{0.92,0,0.59,0.15}
\definecolor{verdes}{cmyk}{0.92,0,0.59,0.4}
\definecolor{verdino}{cmyk}{0.12,0,0.09,0.05}
\definecolor{giallo}{cmyk}{0,0,1,0}
\definecolor{gialloverde}{cmyk}{0.44,0,0.74,0}

\newcommand{\riga}[1]{\noalign{\hbox{\parbox{\textwidth}{#1}}}\nonumber}

\def\plb#1#2#3{{Phys. Lett. }{B #1} (#2) #3}

\font\tenrsfs=rsfs10 at 12pt
\font\sevenrsfs=rsfs7
\font\fiversfs=rsfs5
\newfam\rsfsfam
\textfont\rsfsfam=\tenrsfs
\scriptfont\rsfsfam=\sevenrsfs
\scriptscriptfont\rsfsfam=\fiversfs
\def\mathscr#1{{\fam\rsfsfam\relax#1}}

\newcommand{\be}{\begin{equation}}
\newcommand{\ee}{\end{equation}}
\newcommand{\beq}{\begin{equation}}
\newcommand{\eeq}{\end{equation}}

\def\shat{\ifmmode \hat{s}\else $\hat{s}$\fi}
\def\gp2{{g'}^2}
\def\g2{g^2}
\def\g32{g_s^2}

\newcommand{\newc}{\newcommand}

\newc{\gsim}{\lower.7ex\hbox{$\;\stackrel{\textstyle>}{\sim}\;$}}
\newc{\lsim}{\lower.7ex\hbox{$\;\stackrel{\textstyle<}{\sim}\;$}}
\newc{\ie}{{\it i.e.}}
\newc{\etal}{{\it et al.}}
\newc{\mev}{\hbox{\rm\,MeV}}
\newc{\tev}{\hbox{\rm\,TeV}}
\newc{\xpb}{\hbox{\rm\, pb}}
\newc{\xfb}{\hbox{\rm\, fb}}

\newc{\G}{{\cal G}}
\newc{\h}{{\cal H}}
\newc{\D}{{\cal D}}
\newc{\E}{{\cal E}}

\newc{\mtop}{M_t}
\newc{\mbot}{m_b}
\newc{\mz}{M_Z}
\newc{\mw}{M_W}
\newc{\alphasmz}{\alpha_s(M_Z)}
\newc{\swsq}{\sin^2\theta_W}
\newc{\cwsq}{\cos^2\theta_W}
\newc{\tw}{\tan\theta_W}
\newc{\cw}{\cos\theta_W}
\newc{\sw}{\sin\theta_W}
\newc{\BR}{\hbox{\rm BR}}
\newc{\zbb}{Z\to b\bar}
\newc{\Gb}{\Gamma (Z\to b\bar b)}
\newc{\Gh}{\Gamma (Z\to \hbox{\rm hadrons})}
\newc{\sgn}{\mbox{sgn}}

\def\eq#1{eq.~(\ref{#1})}
\def\fig#1{fig.~\ref{#1}}

\newcounter{mysubequation}[equation]

\newcommand{\GeV}{\,\mathrm{GeV}}
\newcommand{\MeV}{\,\mathrm{MeV}}

\def\dm2{\delta m^2}
\def\dv2{\delta v^2}
\def\dl{\delta \lambda}
\def\Veff{V_{\rm eff}}
\def\nn{\nonumber}
\def\beq{\begin{equation}}
\def\eeq{\end{equation}}
\def\bea{\begin{eqnarray}}
\def\eea{\end{eqnarray}}
\def\slashchar#1{\setbox0=\hbox{$#1$}           
   \dimen0=\wd0                                 
   \setbox1=\hbox{/} \dimen1=\wd1               
   \ifdim\dimen0>\dimen1                        
      \rlap{\hbox to \dimen0{\hfil/\hfil}}      
      #1                                        
   \else                                        
      \rlap{\hbox to \dimen1{\hfil$#1$\hfil}}   
      /                                         
   \fi}                                         %
\catcode`@=11
\long\def\@caption#1[#2]#3{\par\addcontentsline{\csname
  ext@#1\endcsname}{#1}{\protect\numberline{\csname
  the#1\endcsname}{\ignorespaces #2}}\begingroup
    \small
    \@parboxrestore
    \@makecaption{\csname fnum@#1\endcsname}{\ignorespaces #3}\par
  \endgroup}
\catcode`@=12

\begin{document}

\baselineskip=18pt

\setcounter{footnote}{0}
\setcounter{figure}{0}
\setcounter{table}{0}

\begin{titlepage}
\begin{flushright}
CERN-PH-TH/2012--134\\
RM3-TH/12-9
\end{flushright}
\vspace{.3in}

\begin{center}
{\LARGE\color{magenta}\bf Higgs mass and vacuum stability\\[5pt]
in the Standard Model at NNLO}
\vspace{0.5cm}\\
{\bf Giuseppe Degrassi$^a$, Stefano Di Vita$^a$,
 Joan Elias-Mir\'o$^b$,  Jos\'e R. Espinosa$^{b,c}$, \\ Gian F. Giudice$^{d}$,  
Gino Isidori$^{d,e}$, 
Alessandro Strumia$^{g,h}$}\\
\vspace{0.5cm}
{\em $(a)$ {Dipartimento di Fisica, Universit\`a di Roma Tre and  INFN Sez. Roma Tre, Roma, Italy}}\\
{\em $(b)$ {IFAE, Universitat Aut\'onoma de Barcelona, 08193 Bellaterra, Barcelona, Spain}}\\
{\em $(c)$ {ICREA, Instituci\`o Catalana de Recerca i Estudis Avan\c{c}ats, Barcelona, Spain}}\\
{\em $(d)$ {CERN, Theory Division, CH--1211 Geneva 23,  Switzerland}}\\
{\it $(e)$ INFN, Laboratori Nazionali di Frascati, Via E.~Fermi 40, Frascati, Italy}\\
{\it $(g)$ Dipartimento di Fisica, Universit{\`a} di Pisa and INFN Sez. Pisa, Pisa, Italy}\\
{\it  $(h)$ National Institute of Chemical Physics and Biophysics, Tallinn, Estonia}\\
\vspace{0.5cm}

\end{center}
\vspace{.8cm}

\centerline{\large\bf Abstract}
\begin{quote}\large
We present the first complete next-to-next-to-leading order analysis
of the Standard Model Higgs potential.  We computed the
two-loop QCD and  Yukawa corrections to the relation
between the Higgs quartic coupling ($\lambda$) and the Higgs mass
($M_h$), reducing the theoretical uncertainty in the determination of the critical value of
$M_h$ for vacuum stability to 1 GeV. While $\lambda$ at the Planck scale is remarkably close to zero,
absolute stability of the Higgs potential is
excluded at 98\%~C.L. for $M_h < 126\GeV$. Possible consequences of the near vanishing of $\lambda$ at the Planck scale,
including speculations about the role of the Higgs field during
inflation, are discussed.
\end{quote}

\bigskip
\bigskip

\end{titlepage}

\section{Introduction}
The value of the Higgs mass ($M_h$) measured 
by present ATLAS and CMS
data~\cite{ATLAS:2012ac,Chatrchyan:2012tx}, $M_h =125.5\pm0.5\GeV$, is intriguing: 
it is quite close to the minimum $M_h$ value
that ensures absolute vacuum stability within the Standard Model (SM)
which, in turn, implies a vanishing Higgs quartic coupling ($\lambda$)
around the Planck scale. In order to assess if the measured Higgs mass
is compatible with such a peculiar condition, a precise computation is
needed.

The study of the stability of the SM vacuum has a long history~\cite{R1,R2,R3}
 (see also \cite{R4,R5,R6} and references therein).
The state-of-the-art analyses before the latest
LHC data were done at the next-to-leading order (NLO)
level~\cite{CEQ,IRS,BdCE,Isidori:2007vm,ArkaniHamed:2008ym,Bezrukov:2009db,Hall:2009nd,EEGHR}.  This is
based on two-loop renorma\-lization-group (RG) equations, one-loop
threshold corrections at the electroweak scale (possibly improved with
two-loop terms in the case of pure QCD corrections), and one-loop
improved effective potential (see~\cite{EliasMiro:2011aa} for a
numerically updated analysis).

With this paper all the ingredients necessary for a complete
next-to-next-to-leading order (NNLO) analysis in the strong, top Yukawa and Higgs quartic
couplings become available.  In particular, complete
three-loop beta functions for all the SM gauge couplings have been
presented in~\cite{betagauge3}, while the leading three-loop terms in
the RG evolution of $\lambda$, the top Yukawa coupling ($y_t$) and the
Higgs anomalous dimension have been computed
in~\cite{Chetyrkin:2012rz}.  However, as pointed out
in~\cite{EliasMiro:2011aa}, the most important missing NNLO piece for
the vacuum stability analysis are the two-loop threshold corrections
to $\lambda$ at the weak scale due to QCD and top Yukawa interactions,
because such couplings are sizable at low energy.  The calculation of
such terms is presented in this work.

\begin{figure}[t]
$$\includegraphics[width=0.47\textwidth]{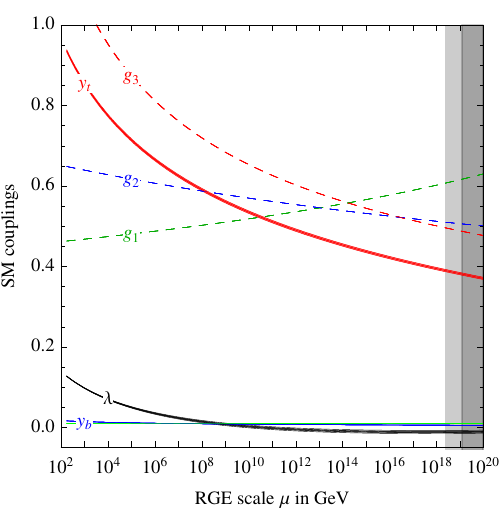}\qquad
\includegraphics[width=0.49\textwidth]{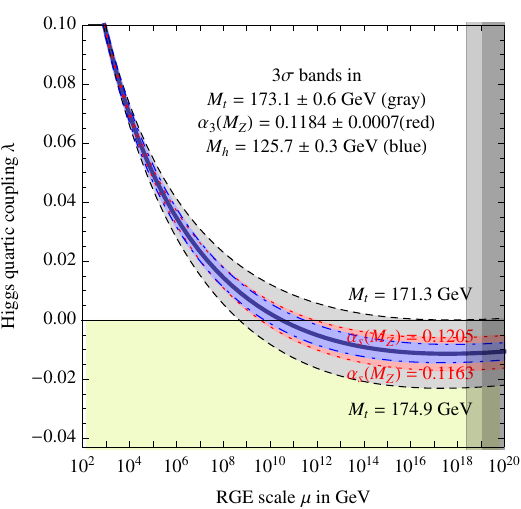}$$
\caption{\em {\bf Left}:
SM RG evolution of the gauge couplings 
$g_1=\sqrt{5/3}g', ~g_2=g, ~g_3=g_s$, of the top and bottom Yukawa couplings ($y_t,y_b$),
and of the Higgs quartic coupling $\lambda$.
All couplings are defined in the  $\overline{\rm MS}$ scheme.
The thickness indicates the $\pm1\sigma$ uncertainty.
{\bf Right}: RG evolution of $\lambda$ varying 
$M_t$, $M_h$ and $\alpha_{\rm s}$ by $\pm 3\sigma$.}
\label{fig:run1} 
\end{figure}

The relation that connects $\lambda$ to the Higgs mass and to the the Fermi 
coupling  ($G_\mu$) can be written as
\be
\lambda (\mu) =\frac{G_\mu M_h^2}{\sqrt{2}} + \Delta \lambda(\mu)~,
\label{eq:defDelta}
\ee 
where $\Delta \lambda(\mu)$ denotes the sizable threshold
corrections arising beyond the tree level.  Given the rapid variation
of $\lambda$ around the weak scale (see \fig{fig:run1}), these
corrections play a significant role in determining the evolution of
$\lambda$ up to high energies.  Computing $\Delta \lambda(\mu)$ at the
one loop level, using two-loop beta functions for all the SM
couplings, and varying the low-energy matching scale between $M_t/2$
and $2M_t$, leads to a $\pm 2$~GeV error on
$M_h$~\cite{EliasMiro:2011aa}.  The NNLO finite terms that we compute
here allow us to reduce this error down to $\pm 0.7\GeV$.  While this
work was in progress an independent calculation of the two-loop
Yukawa-QCD contributions to $\Delta \lambda(\mu)$ has
appeared~\cite{Bezrukov:2012sa}.  Our result agrees with the one in
ref.~\cite{Bezrukov:2012sa} for these contributions.  However, our
analysis includes also the two-loop terms coming from the  Yukawa
sector and can be considered the first complete NNLO evaluation of
$\Delta \lambda(\mu)$.  We stress that both these two-loop terms are
needed to match the sizable two-loop scale dependence of $\lambda$
around the weak scale, caused by the $-32y_t^4 g_s^2+30 y_t^6$
terms in its beta function.  As a result of this improved
determination of $\Delta \lambda(\mu)$, we are able to obtain a
significant reduction of the theoretical error on $M_h$ compared to
previous works.

Putting all the NNLO ingredients together, we estimate an overall
theory error on $M_h$ of $\pm 1.0\GeV$ (see
section~\ref{sec:general}).  Our final results for the condition of
absolute stability up to the Planck scale is 
\beq M_h~[{\rm GeV}]
> 129.4 + 1.4 \left( \frac{M_t~[{\rm GeV}] -173.1}{0.7} \right)
-0.5\left( \frac{\alpha_s(M_Z)-0.1184}{0.0007}\right) \pm 1.0_{\rm th}\ .
\label{stability}
\eeq
Combining in quadrature
the theoretical uncertainty with the experimental errors on $M_t$ and
$\alpha_s$ we get
\beq M_h > 129.4 \pm 1.8~\GeV.
\eeq 
From this result we conclude that vacuum stability of the
SM up to the Planck scale is excluded at $2\sigma$ (98\% C.L.~one sided) for $M_h < 126\GeV$. 

Although the central values of Higgs and top masses do not favor a
scenario with a vanishing Higgs self coupling at the Planck scale
($M_{\rm Pl}$) --- a possibility originally proposed in
ref.~\cite{Nielsen} and discussed more recently in
ref.~\cite{BdCE,Isidori:2007vm,Shaposhnikov:2009pv,Bezrukov:2009db,
Holthausen:2011aa}
--- the smallness of $\lambda$ around $M_{\rm Pl}$ is quite remarkable
(see \fig{fig:run1}). Motivated by this observation, we have
explored in more detail the robustness of the predictions for $M_h$
and $M_t$ assuming special boundary conditions on $\lambda$ and its
beta function around $M_{\rm Pl}$, as advocated in
\cite{Shaposhnikov:2009pv}.  We have also critically examined
scenarios where the Higgs field plays a role during inflation.
This could happen because of a non-minimal coupling to gravity that
flattens the SM potential close to $M_{\rm
  Pl}$~\cite{Bezrukov:2009db}, provided $\lambda(M_{\rm Pl})$ is
positive.  Alternatively, the Higgs field could have caused inflation while it was trapped in a second unstable minimum of the potential that
appears near $M_{\rm Pl}$ if $\lambda(M_{\rm Pl})$ is positive and very
close to zero~\cite{Isidori:2007vm}, provided some non-SM mechanism is
introduced to exit inflation~\cite{Masina:2011aa,Masina:2012yd}.
Beside some technical problems, these frameworks could become viable
only if $M_h$ satisfies the stability condition in
\eq{stability}. We therefore conclude that both these
possibilities are not favored by present data, unless $M_t$ is below about 172 GeV or new-physics threshold corrections at the high scale modify the shape of the SM potential.

The paper is organized as follows: in section~\ref{sec:Delta} we
present the calculation of the threshold correction $\Delta \lambda (\mu)$.
The numerical results for the condition of vacuum stability
and, more generally, for the structure of the SM Higgs potential up to
very high field values are discussed in section~\ref{sec:general}. The
implications of these results for Planck scale physics are discussed
in section~\ref{sec:num}.  The results are summarized in the
conclusions. We include also an appendix in which a ready-to-use expression
for the two-loop effective potential is presented.

\section{The two-loop threshold correction to $\lambda(\mu)$}
\label{sec:Delta}
In this section we present our main new result, namely the calculation
of the two-loop contribution to $\Delta \lambda (\mu)$. We first
obtain the $y_t^6$ and $y_t^4 g_s^2$ terms from the calculation of the
Higgs mass via the effective potential. Then, we present the full
result for the two-loop QCD and  Yukawa contribution to
$\Delta \lambda (\mu)$ in the SM with the electroweak gauge couplings
switched off (the so-called gauge-less limit).

\subsection{Two-loop corrected Higgs mass from the effective potential}
We write the SM potential for the Higgs doublet $H$ in the usual way:
\be \displaystyle V = - m^2 |H|^2 + \lambda |H|^4\qquad
 H =\begin{pmatrix} G^+\cr    (v+h+i G^0)/\sqrt{2}
 \end{pmatrix}\ ,
\label{HiggsPot}
\ee
so that,  up to negligible width effects, the pole Higgs mass $M_h$ is the
solution of the pole equation 
\be M_h^2 = -m^2+3 \lambda v^2 +
\Pi_{hh}(M_h^2)\ , 
\ee 
where $m^2$, $\lambda$ and $v$ are $\overline{\rm MS}$ renormalized
quantities and $\Pi_{hh}(p^2)$ is the Higgs self-energy (two-point)
function, with external four-momentum $p$.  We rewrite this equation
as 
\bea M_h^2 &= &\left[-m^2+3 \lambda v^2 + \Pi_{hh}(0)\right] +
\left[\Pi_{hh}(M_h^2)-\Pi_{hh}(0)\right]\nonumber\\ &=&
     {\left[M_h^2\right]}_V+\Delta\Pi_{hh}(M_h^2)\ .
\label{splitMh2}
\eea
This step is convenient because the last term (which is
computationally challenging) only gives corrections suppressed by the
small Higgs coupling, in view of the smallness of $M_h^2 = 2\lambda
v^2$ at tree level.  The first dominant term can be expressed in term
of derivatives of the effective potential, $\Veff$.  Writing the effective
potential  as a sum of the tree-level part $V_0$ plus radiative
corrections $\Delta V$
\be 
\Veff = -\frac{m^2}{2} h^2 + \frac{\lambda}{4}
h^4 +\Delta V\ , 
\ee one finds 
\be {\left[M_h^2\right]}_V=
\left. \frac{\partial^2 \Veff}{(\partial h)^2}\right|_{h= v}\ ,
\ee
where $v$ is the $h$ vev at the
minimum of the effective potential, determined by the minimization condition
\be
\left.\frac{\partial \Veff}{\partial h}\right|_{h=v}=
\left[ -m^2 h + \lambda h^3 +
\frac{\partial \Delta V}{\partial h}\right]_{h= v}\ .
\ee
As usual, it is convenient to consider $m^2$ as a free parameter fixed in 
terms of $v$ by the above equation, arriving at 
\be
{\left[M_h^2\right]}_V=\left[2\lambda v^2 -
\frac{1}{h}\frac{\partial \Delta V}{\partial h}
+\frac{\partial^2 \Delta V}{(\partial h)^2}\right]_{h= v}\ .
\ee
Defining the operator ${\cal D}_m^2$ as\footnote{Notice that the term in 
${\cal D}_m^2$ linear in field-derivatives automatically 
takes into account the cancellation of $h$-tadpoles (or alternatively, the 
minimization condition to get the right $v$). 
}
\be
{\cal D}_m^2 = \left[-\frac{1}{h}\frac{\partial}{\partial h}+
\frac{\partial^2}{(\partial h)^2}\right]_{h=v}\ ,
\ee
and noting that $2\lambda v^2 = {\cal D}_m^2V_0$, we can simply 
write ${\left[M_h^2\right]}_V={\cal D}_m^2 \Veff $, obtaining the 
following form for the Higgs mass:
\be
M_h^2 = {\cal D}_m^2 \Veff + \Delta \Pi_{hh}(M_h^2)\ .
\label{masterform}
\ee
It gives the Higgs mass squared as the sum of two terms. The first is 
the Higgs mass obtained from the potential; this is not the complete pole 
Higgs mass and must be corrected for nonzero external momentum effects,
which are taken care of by the last term, $\Delta \Pi_{hh}(M_h^2)$.
It is a straightforward exercise to verify that this expression for pole mass is independent
of the renormalization scale $\mu$. In particular, one can easily prove that 
\bea
\frac{d }{d \ln \mu}{\left[M_h^2\right]}_V &=& 
-2\gamma {\left[M_h^2\right]}_V\ ,\\
\frac{d }{d \ln \mu}\Delta \Pi_{hh}(M_h^2)&=& 2\gamma 
\left[M_h^2-\Delta \Pi_{hh}(M_h^2)\right]\ ,
\eea
where $\gamma$ is the Higgs anomalous dimension, describing its wave-function 
renormalization, $\gamma \equiv d\ln h/d\ln \mu$. 

Using \eq{masterform}  and the one-loop result for $\Veff$ in 
\eq{CW} of the appendix one obtains the one-loop
Higgs mass correction.  The explicit one-loop result for the pole mass is
\be
\label{oneloopMh2}
M_h^2=
2  \lambda v^2 +\delta_1 M_h^2\ ,
\ee
with
\bea
\delta_1 M_h^2 &=&
\frac{1}{(4\pi)^2} \left\{3y_t^2(4m_t^2-M_h^2)B_0(m_t,m_t,M_h)
+6\lambda^2 v^2(3\ell_h-6+\pi\sqrt{3})\right.\nonumber\\
&& -\frac{v^2}{4}
(3g^4-8\lambda g^2+16\lambda^2)B_0(m_W,m_W,M_h) \nn \\
&& -\frac{v^2}{8}(3G^4-8\lambda G^2+16\lambda^2)B_0(m_Z,m_Z,M_h)
\nonumber\\
&& + 2m_W^2\left[g^2-2\lambda(\ell_W-1)\right]+m_Z^2
\left[G^2-2\lambda(\ell_Z-1)\right]\left\}\frac{}{}\right. ,
\label{d1Mh}
\eea
where $G^2 = g^2 +g^{\prime \, 2}$. All parameters on the right-hand 
side (including $v$) are
$\overline{\rm MS}$ running parameters (with the exception of $M_h^2$,
which appears through the external momentum dependence of the Higgs
self-energy). As \eq{CW} was obtained in the Landau gauge, 
$v$ in \eq{d1Mh} represents the gauge and scale-dependent vacuum 
expectation value of the Higgs  field   as computed in the Landau gauge. 
Similarly the  $\Delta \Pi_{hh}(M_h^2)$ contribution in that equation is
computed in the Landau gauge.  In \eq{d1Mh}
\be
B_0(m_a,m_b,m_c)\equiv -
\int_0^1 \ln\frac{ (1-x)m_a^2+x m_b^2 -x(1-x) m_c^2-i \epsilon}{\mu^2} dx\ ,
\label{Bfun}
\ee 
and $\ell_x\equiv\ln(m_x^2/\mu^2)$, with $m_x$
the running mass for particle $x$ ($m_t \equiv y_t v/\sqrt{2}$).  One
can explicitly check, using the RGEs for these parameters, that this
expression for $M_h^2$ is indeed scale-independent at one-loop order.

Neglecting gauge couplings and setting $M_h^2=2\lambda v^2$ in the
one-loop terms, one obtains the approximate expression \be \delta_1
M_h^2 \simeq \frac{2y_t^2v^2}{(4\pi)^2} \left[ \lambda(2+3\ell_t)-3
  y_t^2 \ell_t \right]\ .  \ee

To compute \eq{masterform} at the two-loop level one can use the
two-loop effective potential \cite{V2} to calculate
${\left[M_h^2\right]}_V$ and the general results for two-loop scalar
self-energies in \cite{Martinself} (supplemented by the results on
two-loop momentum integrals of \cite{Martinint}) to calculate
$\Delta\Pi_{hh}(M_h^2)$. If we only keep the leading two-loop
corrections to $M_h^2$ proportional to $y_t^6$, $y_t^4 g_s^2$,
dropping all subleading terms that depend on the electroweak gauge
couplings or $\lambda$, our task is simplified dramatically. First, in
the two-loop effective potential we only have to consider the diagrams
depicted in \fig{fig:V2}. Their contribution can be extracted from the expressions for $V_Y$ and $V_{FV}$ in the appendix. 
\begin{figure}[tb]
\includegraphics[width=1\textwidth]{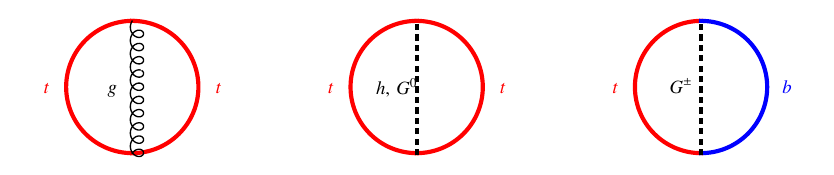}
\caption{\em Two-loop vacuum diagrams that give the dominant contribution (depending only on $g_s$ and $y_t$) to the SM
two-loop effective potential. }\label{fig:V2} 
\end{figure}
Second, in the two-loop term $\Delta \Pi^{(2)}_{hh}(M_h^2)$ we can
substitute the tree-level value $M_h^2=2\lambda v^2$, so that \be
\Delta \Pi^{(2)}_{hh}(M_h^2) \simeq \Pi^{(2)}_{hh}(2\lambda
v^2)-\Pi^{(2)}_{hh}(0)\ .  \ee It is then clear that the two-loop
contributions coming from that term are proportional to $\lambda$ and
are therefore subdominant.  In this section we neglect
$\Delta \Pi^{(2)}_{hh}(M_h^2)$ completely.

\medskip

To find the expression for the Higgs mass at two-loop precision, we
must also take into account that $M_h^2$ has to be evaluated with
one-loop precision in the argument of the one-loop term $\Delta
\Pi^{(1)}_{hh}(M_h^2)$.  Putting together all these pieces, keeping only
the $y_t^6$ and $y_t^4 g_s^2$ terms, we arrive
at the following two-loop correction to \eq{oneloopMh2}:
\beq
\delta_2 M_h^2 =
\frac{y_t^2  v^2}{(4\pi)^4}   \left[16 g_s^2 y_t^2 (3 \ell_t^2+\ell_t)-
3y_t^4 \left(9 \ell_t^2-3 \ell_t+2 +\frac{\pi ^2}{3}\right)  
\right]
\ .
\label{d2Mh}
\eeq

\medskip

The expression for $M_h$ as a function of $\lambda$ can be inverted to obtain 
$\lambda(\mu)$ as a function of the pole Higgs mass $M_h$. To express 
$\lambda(\mu)$
in terms of physical quantities (\ie\  $G_\mu$ and the pole masses $M_Z$, $M_W$, and $M_t$) the relations between physical and $\overline{\rm MS}$ 
parameters are needed. At the level of accuracy we are working only the 
relation between the $y_t(\mu)$ and $M_t$ and the one between $v(\mu)$ and
$G_\mu$ are required. They are given by:
\bea
y_t^2(\mu) &= &2\sqrt{2}G_\mu M_t^2\left[
1+\frac{8}{3}\frac{1}{(4\pi)^2}  g_s^2(3L_T-4)+
\frac{1}{(4\pi)^2}  \sqrt{2}G_\mu M_t^2(-9L_T+11)\right] , \label{htq} \\
v^2(\mu)&=&
\frac{1}{\sqrt{2} G_\mu}+\left.\frac{1}{(4\pi)^2}  
\right[3 M_t^2 (2 L_T-1)+ M_W^2 (5-6 L_W)+\frac{1}{2}M_Z^2 (5-6 L_Z)\nonumber\\
&&\left.
+\frac{3 M_Z^2 M_W^2 }{4(M_Z^2-M_W^2)}(L_Z-L_W)-
\frac{1}{2}M_h^2 - \frac{3M_W^2M_h^2}{M_W^2- M_h^2} (L_W-L_H)\right]\ ,
\label{vGmu}
\eea
where $L_X=\ln(M_x^2/\mu^2)$, with masses in capital letters 
denoting pole masses. 

\medskip

We find:
\be
\label{SZext}
\lambda(\mu)=\frac{G_\mu M_h^2}{\sqrt{2}}+\lambda^{(1)}(\mu)+
\lambda^{(2)}(\mu)\ ,
\ee
with
\be
\lambda^{(2)}(\mu) =
\lambda^{(2)}_{\rm QCD, lead.}(\mu)+\lambda^{(2)}_{\rm Yuk, lead.}(\mu) + \ldots\ ,
\ee
where the ellipsis stands for the subleading terms neglected in this section.
The known one loop term is
\bea
\lambda^{(1)}(\mu) &=&
\frac{1}{2} G_\mu^2 \frac{1}{(4\pi)^2}  
\left\{\frac{6 (L_H-L_W) M_h^6}{M_h^2-M_W^2}-8 \left(2 M_W^4+M_Z^4\right)-
2 (-3+6 L_T) M_h^2 M_t^2\right.\nonumber\\
&&+M_h^4 \left(19-15 L_H+6 L_W-3 \sqrt{3} \pi \right)+
12 (M_h^2-4 M_t^2) M_t^2 B_0(M_t,M_t,M_h)\nonumber\\
&&+2 \left(M_h^4-4 M_h^2 M_W^2+12 M_W^4\right) B_0(M_W,M_W,M_h)\nn \\
&& + \left(M_h^4-4 M_h^2 M_Z^2+12 M_Z^4\right) B_0(M_Z,M_Z,M_h)\nonumber\\
&&\left.
+ M_h^2 \left[2(8L_W-7) M_W^2+(8L_Z-7) M_Z^2-
\frac{6  M_Z^2 M_W^2}{M_Z^2-M_W^2}(L_Z-L_W)\right]\right\}\ ,
\eea
and the leading two loop QCD and Yukawa terms are
\bea
\lambda^{(2)}_{\rm QCD, lead.}(\mu)
&=& \frac{G_\mu^2 M_t^4 }{(4\pi)^4}64   g_s^2(\mu) 
\left(-4-6 L_T+3 L_T^2\right)\ , \\
\lambda^{(2)}_{\rm Yuk, lead.}(\mu)
&=& \frac{8\sqrt{2}G_\mu^3 M_t^6 }{(4\pi)^4} 
\left(30+\pi ^2+36 L_T-45 L_T^2\right) \ .
\eea
The above expression for $\lambda(\mu)$ has the correct dependence on the
renormalization scale $\mu$, so that both sides of (\ref{SZext}) evolve
with $\mu$ in the same way to the order we work.

\subsection{Two-loop contribution to $\Delta \lambda (\mu)$ in the gauge-less 
SM} 
In this section we go beyond the leading $y_t^6$ and $y_t^4 g_s^2$
contributions to $\Delta \lambda (\mu)$ discussed in the previous
section and derive the QCD and Yukawa corrections in the SM with the
electroweak gauge couplings switched off. We first discuss the
two-loop renormalization of the Higgs sector of the SM and then derive
the 2-loop relation between $\lambda(\mu)$ and the physical parameters
$G_\mu, M_t$, and $M_h$. In our derivation we follow closely
ref.\cite{SZ}.

We start from the unrenormalized Higgs potential, \eq{HiggsPot} written
in terms of bare quantities, and set
$m^2 \to m_r^2 - \dm2,\: 
\lambda \to \lambda_r- \dl, \: v \to v_r -\delta v$. Then, assuming 
$\dm2,\: \dl,\: \delta v$ of ${\cal O}
(\alpha)$,  we obtain
\beq 
V= V_{r} - \delta V\ ,
\eeq 
where, putting $m_r^2 = \lambda_r v_r^2$, 
\bea  
 V_{r} & = & \lambda_r \left[G^+ G^- \left( G^+ G^- + h^2 + G_0 \right)^2 + 
                   \frac14 \left( h^2 + G_0^2\right)^2 \right] \nn \\
           & & +\, \lambda_r v_r \,h \left[ h^2 + G_0^2 + 2\, G^+ G^-\right] +
            \frac12 M_h^2 \,h^2\ ,
\label{Rpot}
\eea
with $M_h = 2\, \lambda_r v_r^2$, while, up to two-loop terms,
\bea
\delta V &=& \dl \left[G^+ G^- \left( G^+ G^- + h^2 + G_0 \right)^2 + 
                   \frac14 \left( h^2 + G_0^2\right)^2 \right] \nn \\
        & & + \left[ \lambda_r \left( \frac{\dv2}{2\,v_r} + 
               \frac{(\dv2)^2}{8\,v_r^3} \right) + v_r \,\dl 
            \left(1 -   \frac{\dv2}{2\,v_r^2}\right) \right]
           h \left[ h^2 + G_0^2 + 2\, G^+ G^-\right] \nn \\
        & & + \delta \tau \left( \frac12 G_0^2 + G^+ G^- \right) +
         \frac12 \delta M_h^2  h^2 + v_r \,\delta \tau 
        \left( 1 - \frac{\dv2}{2\, v_r^2} \right) \, h~.
\label{deltaV} 
\eea
In \eq{deltaV}
\bea
\delta M_h^2 &=& 3 \left[ \lambda_r \dv2 + v_r^2 \dl 
 \left(   1 - \frac{\dv2}{v_r^2} \right) \right] - \dm2  
\label{dmh}\ ,\\
\delta \tau &=&  \lambda_r \dv2 + v_r^2 \dl 
 \left(   1 - \frac{\dv2}{v_r^2} \right) - \dm2 \ ,
\label{dtau}
\eea
and $\dv2$ is related to $\delta v$ through 
$\sqrt{v_r^2 - \dv2} = v_r - \delta v$.

Following ref.\cite{SZ} we require the cancellation of the tadpole contribution
by setting
\beq
\delta \tau 
        \left( 1 - \frac{\dv2}{2\, v_r^2} \right) = -\frac{T}{v_r}\ ,
\label{tadpole}
\eeq
where $i T$ is the sum of the tadpole diagrams with the external leg extracted.
We identify $M_h^2$ in $V_r$ with the
on-shell Higgs mass  leading to the condition
\beq 
\delta M_h^2 = {\rm Re}\, \Pi_{hh}(M_h^2)\ ,
\label{mass}
\eeq
where the $ \Pi_{hh}(M_h^2)$ in the above equation includes only the
contribution of the self-energy diagrams because \eq{tadpole} is enforced.
A third condition can be obtained by looking
directly at the muon-decay process. At the two-loop level
we can write
\beq
\frac{G_\mu}{\sqrt{2}} = \frac1{2 v_0^2} \left\{1 - \frac{A_{WW}}{M_{W_{0}}^2}
+ V_W + M_{W_{0}}^2 B_W 
 + \left(\frac{A_{WW}}{M_{W}^2}\right)^2 - \frac{A_{WW} V_W}{M_W^2} 
\right\} \ ,
\label{Gmu}
\eeq 
where $v_0$ is the unrenormalized vacuum, $A_{WW} \equiv A_{WW} (0)$ is the 
$W$ self-energy
evaluated at zero external momenta, $V_W$ and $B_W$ are the relevant vertex
and box contributions in the $\mu$-decay process and $M_{W_{0}}$ is the
unrenormalized $W$ mass. Performing the shifts $v_0^2 \to v_r^2 - \dv2$,
$M_{W_{0}} \to M^2_{W} - \delta M^2_W$, where $\delta M^2_W = 
{\rm Re} A_{WW} (M_W^2)$,  and working at the two-loop level
we arrive at
\beq
 v_r^2   =  \frac1{\sqrt{2}  \,G_\mu} -
\frac1{\sqrt{2}  \,G_\mu} \left\{ \frac{A_{WW}}{M_{W}^2}
- E + \frac{ A_{WW}\, \delta M_{W}^2}{M_W^4}
 - \left(\frac{A_{WW}}{M_{W}^2}\right)^2 + \frac{A_{WW} V_W}{M_W^2} +
\delta M_W^2 B_W \right\} + \dv2\ ,
\label{Vacuum}
\eeq
where $ E \equiv V_W + M_{W}^2 B_W$.  
 We identify the renormalized vacuum by $v_r^2 = 1/(\sqrt{2}  \,G_\mu)$, then
$\dv2$ is defined to cancel the contribution of the curly bracket in 
\eq{Vacuum}.

Our choice of renormalization conditions implies that the renormalized
quartic Higgs coupling is set equal to
\beq
\lambda_r = \frac{G_\mu}{\sqrt{2}} M_h^2\ ,
\label{lambda}
\eeq
while eqs.(\ref{dmh}--\ref{Vacuum}) can be used to obtain the correction $\dl$.
Writing
\beq
\dl = \dl^{(1)} + \dl^{(2)}~,
\eeq
where the superscript indicates the loop order, we have
\bea
\dl^{(1)} & = & - \frac{G_\mu}{\sqrt{2}} M_h^2 \left\{
\frac{A_{WW}^{(1)}}{M_{W}^2} - E^{(1)} -
\frac1{M_h^2} \left[{\rm Re}\, \Pi_{hh}^{(1)}(M_h^2) + 
\frac{T^{(1)}}{v_r} \right] \right\}  
\label{deltalh1} \\
\dl^{(2)} & = & - \frac{G_\mu}{\sqrt{2}} M_h^2 \left\{
\frac{A_{WW}^{(2)}}{M_{W}^2} - E^{(2)}  -
\frac1{M_h^2} \left[{\rm Re}\, \Pi_{hh}^{(2)}(M_h^2) + \frac{T^{(2)}}{v_r} \right] 
\right. \nn \\
&+ & \left( \frac{A_{WW}^{(1)}}{M_{W}^2} - E^{(1)} \right)
     \left( \frac{A_{WW}^{(1)}}{M_{W}^2} - E^{(1)} -
\frac1{M_h^2} \left[{\rm Re}\, \Pi_{hh}^{(1)}(M_h^2) + \frac32 
\frac{T^{(1)}}{v_r}  \right] \right) \nn \\
& +& \left. \frac{ A_{WW}^{(1)}\, \delta^{(1)} M_{W}^2}{M_W^4}
 - \left(\frac{A_{WW}^{(1)}}{M_{W}^2}\right)^2 + 
\frac{A_{WW}^{(1)} V^{(1)}_W}{M_W^2} + \delta^{(1)} M_W^2 B_W^{(1)} \right\}~.
\label{deltalh2}  
\eea

The connection between $\lambda_r$, as defined in \eq{lambda}, and
$\lambda(\mu)$ can be easily derived using
\beq
\lambda_r - \dl = \lambda(\mu) - \delta \hat{\lambda}\ ,
\label{rel}
\eeq
or
\beq
\lambda(\mu) =  \frac{G_\mu}{\sqrt{2}} M_h^2  - \dl + \delta \hat{\lambda}~.
\label{rel2}
\eeq
In eqs.~(\ref{rel})--(\ref{rel2})  $\delta \hat{\lambda}$ is the 
counterterm associated to $\lambda(\mu)$, \ie\ the counterterm that 
subtracts only the terms proportional to powers of 
$1/\epsilon$ and $\gamma -\ln (4 \pi)$
in dimensional regularization, with $d= 4- 2\, \epsilon$ being the dimension of 
space-time. Concerning  the structure of the $1/\epsilon$ poles
in $\dl$ and $\delta \hat{\lambda}$, one  notices that it should be  
identical once the poles in $\dl$ are expressed in terms of 
$\overline{\rm MS}$ quantities. Then, after this
operation is performed, a finite $\lambda(\mu)$ is obtained.

Specializing the above discussion to the two-loop case we have
\beq
\lambda(\mu) = \frac{G_\mu}{\sqrt{2}} M_h^2 - \dl^{(1)}|_{\rm fin} -
                   \dl^{(2)}|_{\rm fin} + \Delta
\label{finlam}
\eeq
from which we identify the one- and two-loop contributions entering \eq{SZext},
\bea
\lambda^{(1)}(\mu) &=& - \dl^{(1)}|_{\rm fin} \label{d1l}~, \\[2mm]
\riga{which reproduces the one-loop result of ref.\cite{SZ} and}\\[-2mm]
\lambda^{(2)}(\mu) &=& - \dl^{(2)}|_{\rm fin} + \Delta \label{d2l} ~.
\eea
In eqs.~(\ref{finlam})--(\ref{d2l})
the subscript `fin' denotes the finite part of the quantity involved 
and $\Delta$ is the two-loop finite contribution that is obtained
when the OS parameters entering the  1/$\epsilon$ pole in  $\dl^{(1)}$ are 
expressed in terms of $\overline{\rm MS}$ quantities, the finite
contribution coming from the $O(\epsilon)$ part of the shifts.  

Differentiating \eq{finlam} with respect to 
$\mu$, the known two-loop beta function for the Higgs quartic coupling is 
recovered. It should be recalled  that the right-hand side of \eq{finlam} is 
expressed in terms of physical quantities, then
the dependence on $\mu$ in that equation is explicit. To obtain the correct
two-loop beta function, one has  first to differentiate with respect to $\mu$
and then to express  the one-loop part in terms of $\overline{\rm MS}$ 
quantities.

The computation of  $\lambda^{(2)}(\mu)$ in the full 
SM is quite cumbersome, see \eq{deltalh2}. However, the calculation can 
be greatly simplified if one considers the gauge-less limit of the SM
in which the electroweak interactions are neglected, \ie\ the gauge couplings
$g$ and $g^\prime$ are set equal to zero. In this limit, \eq{deltalh2} simplifies
to
\bea
\left(\dl^{(2)} - \Delta \right)_{g.l.} & = & 
- \frac{G_\mu}{\sqrt{2}} M_h^2 \left\{
\frac{A_{WW}^{(2)}}{M_{W}^2}   -
\frac1{M_h^2} \left[{\rm Re}\, \Pi_{hh}^{(2)}(M_h^2) + \frac{T^{(2)}}{v_r} \right] 
\right. \nn \\
&+ & \left. \frac{A_{WW}^{(1)}}{M_{W}^2} 
     \left( \frac{A_{WW}^{(1)}}{M_{W}^2} -
\frac1{M_h^2} \left[{\rm Re}\, \Pi_{hh}^{(1)}(M_h^2) + \frac32 
\frac{T^{(1)}}{v_r}  \right] \right) \right\}_{g.l.} - \Delta_{g.l.}\ ,
\label{deltalhgl}  
\eea
where the subscript $g.l.$ means that we have considered in the various
self-energies only diagrams involving the top and bottom quarks, the Higgs
and the Goldstone bosons, the latter with vanishing mass, and the limit
$g,g^\prime \to 0$ is taken.

Using \eq{deltalhgl} we compute  the QCD and the
 Yukawa contribution to
$\lambda^{(2)}(\mu)$. The top Yukawa-QCD contribution, 
$\lambda^{(2)}_{\rm QCD}(\mu)$, is obtained evaluating the relevant diagrams 
via a Taylor series in
$x_{ht}\equiv M_h^2/M_t^2$ up to fourth order
\bea
\lambda^{(2)}_{\rm QCD}(\mu)
&=& \frac{G_\mu^2 M_t^4 }{(4\pi)^4}\,N_c\, C_F\,g_s^2(\mu) \bigg[ 16   
\left(-4-6 L_T+3 L_T^2\right)  \nn \\
&+& \left. x_{ht} \left( 35-\frac{2\,\pi^2}3 +12 L_T- 12 L_T^2\right) 
+  x_{ht}^2 \frac{61}{135} + 
x_{ht}^3 \frac{1223}{6300} + 
x_{ht}^4 \frac{43123}{1323000}\right] \ ,
\label{deltaQCD}
\eea
where $N_c$ and $C_F$ are color factors ($N_c=3, \, C_F=4/3$).
Equation~(\ref{deltaQCD}) shows that the series converges very fast. Our result is 
in agreement  with ref.\cite{Bezrukov:2012sa}, the numerical difference 
between \eq{deltaQCD} and the expression of ref.\cite{Bezrukov:2012sa}
for $M_h \sim 125$ GeV being negligible.

The  Yukawa contribution, $\lambda^{(2)}_{\rm Yuk}(\mu)$, is (neglecting the small bottom Yukawa)
\bea
\lambda^{(2)}_{\rm Yuk}(\mu)
&=& \frac{\sqrt{2} G_\mu^3 M_t^6 }{(4\pi)^4} \Bigg\{
N_c^2 \Bigg[ 16 B_0(M_t,M_t,M_h) (-1+ 2 L_T)  \nn \\
&&~~~~~~~~~~~~~~~~ + 
x_{ht} \left( (1 + 4 B_0(M_t,M_t,M_h) - 2 L_T)(1 - 2 L_T) \right) \Bigg]  \nn \\ 
&&+ N_c 
\left[ 16 + \frac83 \pi^2 + 32 B_0(M_t,M_h,M_t)(1+ 2 L_T)  - 48 L_T +
40 L_T^2 \right. \nn\\
&&~~~~~~~~~- x_{ht} \left( \frac{929}6 +\frac{16}3 \pi^2  
+48 B_0(M_h,M_h,M_h) - 16 L_H \left(1- L_T \right)\right. \nn \\
&&~~~~~~~~~~~~~~~ \left. 
 +B_0(M_t,M_h,M_t) \left( \frac{76}3 + 32 L_T \right) 
+ \frac{190}3 L_T +58 L_T^2 \right) \nn \\
&&~~~~~~~~~ + x_{ht}^2 \left(\frac{17629}{270} +\frac{8}3 \pi^2  
-\frac23 L_H + B_0(M_h,M_h,M_h)\left(27-18 L_T\right) +40 L_T \right. \nn \\
&&~~~~~~~~~~~~~~~ \left. + 10 L_T L_H +12 L_T^2 + 
B_0(M_t,M_h,M_t) \left( \frac{13}3 + 4 L_T \right) \right) \nn\\ 
&&~~~~~~~~~ + x_{ht}^3  \left(\frac{1181}{900} - \frac{ \pi^2}2 +
\frac{61}{30} B_0(M_h,M_h,M_h)  + \frac{59}{90}L_H \right.   \nn \\
&&~~~~~~~~~~~~~~~ \left. \left. - \frac2{35} B_0(M_t,M_h,M_t)-
\frac{68}{63} L_T \right)\right] \nn\\
&&+ x_{ht}^3 \left[ \frac{131}6 \pi^2 + \left( \frac{729}2
-\frac{135}4 \sqrt{3}\, \pi \right) S_2 - 111 L_H + 36 L_H^2 
\right. \nn \\
&&~~~~~~~~~~~~~  \left. \left. +\pi \left( 
\frac{-225 \sqrt{3}}4 + 18 \sqrt{3} L_H \right) + \frac{75 + 72\, \zeta_3}4
\right] \right\} \ ,
\label{deltaYuk}
\eea
where $ B_0(M_h,M_h,M_h) = 2 - L_H -\pi/\sqrt{3}$ and
$S_2 = 4/(9\, \sqrt{3})\,  {\rm Cl}_2(\pi/3) = 0.260434138\ldots$
In \eq{deltaYuk} the terms proportional to $N_c^2$ and $N_c^0$ were
 computed exactly while
the ones proportional to $N_c$ were computed via an asymptotic expansion in
the large top mass up to $x_{ht}^3$ terms exploiting the asymptotic-expansion 
techniques developed in ref.\cite{DS}. The part independent of $N_c$ in \eq{deltaYuk} was computed using the results for the
two-loop on-shell master integrals of ref.\cite{Fleischer:1999hp}.

We end this section by commenting on the size of the terms suppressed by powers of $x_{ht}$ with
respect to the $y_t^4 g_s^2$ and $y_t^6$ contribution in eqs.~(\ref{deltaQCD},
\ref{deltaYuk}). While in the QCD case, \eq{deltaQCD}, the $x_{ht}$
suppressed terms are indeed smaller than the $y_t^4 g_s^2$ contribution,
the same is not true in the  Yukawa case, \eq{deltaYuk}, where 
the $x_{ht}$ terms are actually larger than the $y_t^6$ contribution.

\section{Extrapolating the SM up to the Planck scale}
\label{sec:general}

A full NNLO computation of the Higgs potential requires three main
ingredients: 1)~the two-loop effective potential; 2)~three-loop beta
functions for all the relevant couplings; 3)~two-loop matching
conditions to determine the initial values of the couplings at the
electroweak scale.  As anticipated in the introduction, all these
ingredient are now available for the QCD, Yukawa and Higgs quartic 
couplings. In this section we first discuss the structure of the
two-loop potential and the numerical inputs at the electroweak scale,
and then present the final numerical results for the stability
condition in the $M_h$--$M_t$ plane.

\subsection{The two-loop effective potential}
 
The SM effective potential is known up to two-loops \cite{V2}. Its
explicit form in a ready-to-use expression is given in the
appendix.  
For large field values ($h \gg v$), the potential is very well approximated by its RG-improved tree-level expression,
\be 
V^{\rm tree}_{\rm eff}(h)  =
\frac{\lambda(\mu)}{4} h^4~,
\label{eq:Vefftree}
\ee 
with $\mu = {\cal O}(h)$. For this reason, if we are interested only in
the condition of absolute stability of the potential, we could simply
study the RG evolution of $\lambda$ imposing the condition
$\lambda(\Lambda)\geq 0$ for any value $\Lambda$ up to the Planck
scale (as for instance done in~\cite{IRS}).  Given that $\lambda$
reaches its minimum value before $M_{\rm Pl}$, independently of its
initial condition at the electroweak scale, the minimum Higgs mass
ensuring vacuum stability corresponds to the initial value of
$\lambda$ such that at some scale $\Lambda_0$ 
\be \lambda(\Lambda_0) =
\beta_\lambda (\Lambda_0)= 0~, \qquad \qquad \beta_\lambda =
\frac{d }{d\ln\mu } \lambda(\mu)~.
\label{eq:beta}
\ee
This is indeed the condition analyzed in ref.~\cite{Bezrukov:2012sa}.
In principle, a more accurate determination of the minimal $M_h$ 
ensuring vacuum stability is obtained taking into 
account the full structure of the Higgs potential at the two-loop level. 
In practice, the  determination of $M_h$ obtained by the condition (\ref{eq:beta})
differs by about 0.1~GeV from the one determined by the absolute 
stability of the RG-improved two-loop potential.

In the following we are interested also in analyzing the shape of the
Higgs potential close to the Planck scale and in the scale where the
instability occurs (as a function of $M_h$ and $M_t$). To this purpose, the study of 
the RG evolution of $\lambda$ is not sufficient and the complete structure of the 
effective potential at the two-loop level plays a significant role.
As pointed out in~\cite{CEQ}, one can
always define an effective coupling $\lambda_{\rm eff}(h)$ such that
for $h\gg v$ the two-loop effective potential assumes the form 
\be
V_{\rm eff}(h) = \frac{\lambda_{\rm eff}(h) }{4} h^4~.
\label{eq:Veffsimpl}
\ee
The explicit two-loop result for $\lambda_{\rm eff}(h)$ can be easily obtained from the two-loop potential and is given in the appendix.
We report here the simplified expression obtained when, in the two-loop term,
we take into account only the contributions from the strong and the top Yukawa couplings\footnote{At high scales, the
electroweak gauge couplings $g'$ and $g$ become comparable in size to $y_t$ and $g_s$ (see \fig{fig:run1}), but their contribution to $\lambda_{\rm eff}(h)$ turns out to be numerically small so that \eq{lambdaeff} is a very good approximation.} \cite{CEQ}:
\bea\label{lambdaeff}
\lambda_{\rm eff}(h)&=&e^{4\Gamma(h)}\left\{\lambda(h) + \frac{1}{(4\pi)^2} \sum_p N_p \kappa_p^2\left(r_p -C_p\right)\right.\\
&+&\left.\frac{1}{(4\pi)^4} y_t^4\left[8g_s^2(3r_t^2-8r_t+9)-\frac{3}{2}y_t^2\left(3r_t^2-16r_t+23+\frac{\pi^2}{3}\right)\right]
\right\}\ . \nonumber
\eea
Here all couplings  are evaluated at the  scale determined by the field value ($\mu=h$),  the index $p$ runs over $p$article species, 
$N_p$ counts degrees of freedom (with a minus sign for fermions), the field-dependent mass squared of 
species $p$ is $m_p^2(h)=\mu_p^2+\kappa_p h^2$ and $C_p$ is a constant. The values of $\{N_p,C_p,\mu_p^2,\kappa_p\}$
within the SM are:
\beq\begin{array}{c|ccccc}
p &  t& W & Z & h &\chi\\ \hline
N_p & -12&6&3&1&3\\
C_p &3/2 &5/6 &5/6 & 3/2& 3/2\\
\mu_p^2 &0 &0 &0 & -m^2& -m^2\\
\kappa_p & y_t^2/2&g^2/4&(g^2+{g'}^2)/4&3\lambda&\lambda
\end{array}\eeq
The  factor 
\be
\Gamma(h)\equiv\int_{M_t}^h \gamma(\mu)\, d\ln \mu\ ,
\ee 
where $\gamma\equiv d\ln h/d\ln \mu$ is the Higgs field anomalous dimension, takes into account the wave-function renormalization.
We have also defined $r_p\equiv \ln[\kappa_p e^{2\Gamma(h)}]$.

\begin{figure}[t]
$$\includegraphics[width=0.45\textwidth]{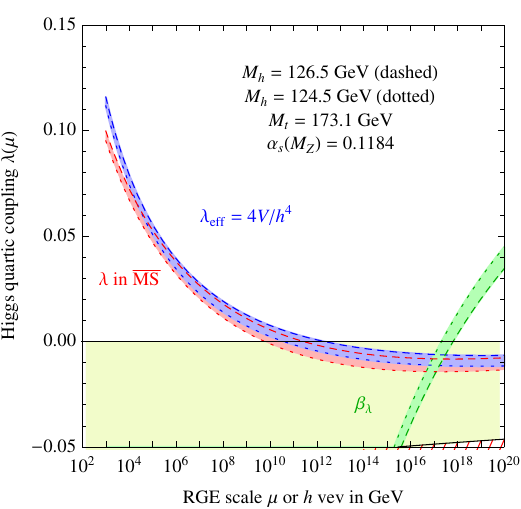}\qquad
\includegraphics[width=0.45\textwidth]{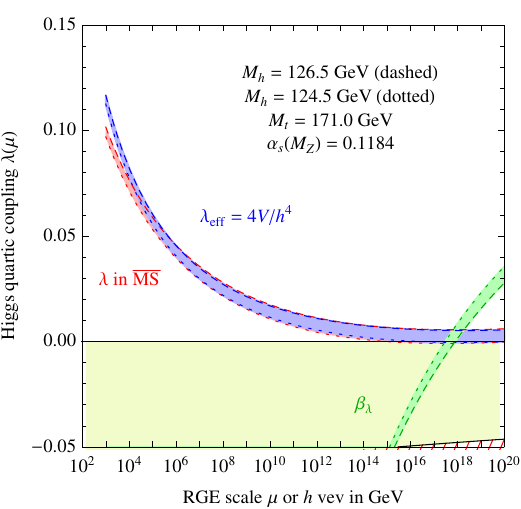}$$
\caption{\em Evolution of the Higgs coupling $\lambda(\mu)$ and its
beta function, \eq{eq:beta}, as a function of the 
renormalization scale, compared to the evolution of the effective coupling 
$\lambda_{\rm eff}(h)$, defined in \eq{eq:Veffsimpl}, as a function of the field value.
 {\bf Left}: curves plotted for the best-fit value of $M_t$. {\bf Right}: curves plotted  for the lower value of $M_t$
that corresponds to $\lambda(M_{\rm Pl})=0$.
\label{fig:run2} }
\end{figure}

The difference $\lambda_{\rm eff}(h)-\lambda(h)$ is positive,  as illustrated in \fig{fig:run2}. 
As a result \cite{CEQ}, at a given field value the potential is more stable than what guessed from the 
naive expectation based on the RG-improved tree-level potential in \eq{eq:Vefftree}, with $\mu=h$.
We finally notice that the difference $\lambda_{\rm eff}(h)-\lambda(h)$ gets suppressed at large 
field values, especially when $\lambda$ reaches its minimum close to the Planck scale.  
This is expected according to the following two observations: 1)~the 
difference between $\lambda_{\rm eff}$  and $\lambda$ can be reabsorbed by a shift in the 
scales at which the two couplings are evaluated, up to finite two-loop corrections;~2) this shift has a small impact at large 
field values given the corresponding vanishing of $\beta_\lambda$  (see \fig{fig:run2}).

\subsection{Inputs at the electroweak scale and threshold corrections}
\label{sec:inputs}
As far as the SM gauge couplings are concerned, we can directly use  results 
in the literature for the couplings in the $\overline{\rm MS}$ scheme. In particular, from a global 
fit of electroweak precision data, performed with the additional input $M_h \approx 125 \GeV$,
the following $\overline{\rm MS}$ values of the electromagnetic coupling and the weak angle
renormalized at $M_Z$ are obtained~\cite{alpha}:
\beq \alpha_{\rm em}^{-1}=127.937 \pm 0.015~,\qquad
\sin^2\theta_{\rm W}= 0.23126 \pm 0.00005~.\eeq
From these we derive
\beq \alpha_2^{-1}(M_Z) = \alpha_{\rm em}^{-1}\sin^2\theta_{\rm W}=
29.587\pm 0.008~,\eeq
\beq \alpha_Y^{-1}(M_Z)=\alpha_{\rm em}^{-1}\cos^2\theta_{\rm W}=98.35\pm 0.013 \ .
\eeq
For the strong coupling we adopt
\beq \alpha_s(M_Z)=0.1184\pm 0.0007~\hbox{\cite{alpha3}}
\ee
such that, including 3 loop RG running up to $M_t$ and matching to the theory with 6 flavors, we get
\beq g_s(M_t)  = 1.1645 +0.0031\left( \frac{\alpha_s(M_Z) - 0.1184}{0.0007}\right)
-0.00046 \left(\frac{M_t}{\GeV}-173.15 \right).\eeq
We determine the  $\overline{\rm MS}$ top-quark Yukawa coupling ($y_t$) starting from the top-quark pole mass ($M_t$)
determined from experiments. Averaging  measurements  from Tevatron and LHC experiments,
\be
M_t = \left\{\begin{array}{ll} 173.2\pm0.9~{\rm GeV} & \hbox{Tevatron~\hbox{\cite{topmass}}} \\
172.6 \pm 0.6\pm 1.2~{\rm GeV} & \hbox{CMS $\mu j$~\cite{MtCMS}}\\
174.5\pm0.6\pm 2.3~{\rm GeV} & \hbox{ATLAS $\ell j$~\cite{MtATLAS}},
\end{array}\right.
\ee
we get 
\beq M_t =(173.1\pm0.7)\GeV~.
\label{expMt}
\eeq 
In order to translate this value into a determination of $y_t$
we apply: 1)  QCD threshold corrections up to $O(\alpha_s^3)$~\cite{Broadhurst:1991fy,Melnikov:2000qh};
2) complete one-loop electroweak corrections from ref.~\cite{Hempfling:1994ar};
3) two-loop ${\cal O}(\alpha \alpha_s)$ corrections from~ref.~\cite{Jegerlehner:2003py}, including the ${\cal O}(\alpha \alpha_s)$ 
terms due to the renormalization of the Fermi coupling (see sect.~\ref{sec:Delta}). 
As a result, we find, for the $\overline{\rm MS}$ top Yukawa coupling renormalized at the top pole mass $M_t$:
\bea
y_t(M_t) &=& 0.93587
+0.00557 \left(\frac{M_t}{\GeV}-173.15 \right) -0.00003  \left(\frac{M_h}{\GeV}-125\right) \nonumber \\
&& -0.00041\left( \frac{\alpha_s(M_Z) - 0.1184}{0.0007}\right) \pm0.00200_{\rm th}~. 
\label{eq:ht_ew}
\eea 
The ${\cal O}(\alpha \alpha_s)$ term, that is the parametrically smallest
correction, is equivalent to a tiny shift in $M_t$ below
$0.1$~GeV. This effect is well below the ${\cal O}(\Lambda_{\rm QCD})$
irreducible non-perturbative uncertainty on the top-quark mass
determined at hadron colliders (see e.g.~ref.~\cite{Hoang:2008xm,R7}),
that is responsible for the theoretical error in \eq{eq:ht_ew}.
More explicitly, we estimate an irreducible theoretical error of $\pm
\Lambda_{\rm QCD}\approx \pm 0.3\GeV$ in $M_t$ from non-perturbative
effects, and an additional uncertainty of $\pm 0.15\GeV$ from missing
${\cal O}(\alpha^4_s)$ threshold corrections.\footnote{~In principle, 
a direct determination of the  $\overline{\rm MS}$ top-quark mass
 at hadron colliders can be obtained from the experimental data on the $\sigma(pp/p\bar p \to t\bar t)$ cross 
section (see ref.~\cite{R7b} and references therein).
At present  this determination leads to a value for $M_t$ which is perfectly consistent with eq.~(\ref{expMt}) 
but has an error four times larger~\cite{R8}.
For completeness and for future reference, we report here the stability condition in eq.~(\ref{stability}) 
as a function of the $\overline{\rm MS}$ top-quark Yukawa coupling, rather than the top-quark pole  mass:
$$
 M_h~[{\rm GeV}]
> 129.4 + 2.0 \left( \frac{y_t(M_t) -0.9356}{0.0054} \right)
-0.35\left( \frac{\alpha_s(M_Z)-0.1184}{0.0007}\right) \pm 1.0_{\rm th}\ .
$$}

Next, applying the threshold corrections discussed in
section~\ref{sec:Delta}, we determine the following value for the
Higgs self coupling in the $\overline{\rm MS}$ scheme renormalized at
the pole top mass: 
\be \lambda(M_t) = 0.12577+0.00205
\left(\frac{M_h}{\GeV}-125\right)-0.00004\left(\frac{M_t}{\GeV}-173.15\right)\pm
0.00140_{\rm th}~.  \ee 
The residual theoretical uncertainty, that is
equivalent to an error of $\pm 0.7$~GeV in $M_h$, has been estimated
varying the low-energy matching scale for $\lambda$ between $M_Z$ and
$2M_t$.

For completeness, we also include in the one- and two-loop RG equation
the contributions of the small bottom and tau Yukawa couplings, as
computed from the $\overline{\rm MS}$ $b$-quark mass,
$m_b(m_b)=4.2\GeV$, and from $M_\tau=1.777\GeV$.

\subsection{Phase diagram of the SM}

\begin{figure}[t]
$$\includegraphics[width=0.45\textwidth]{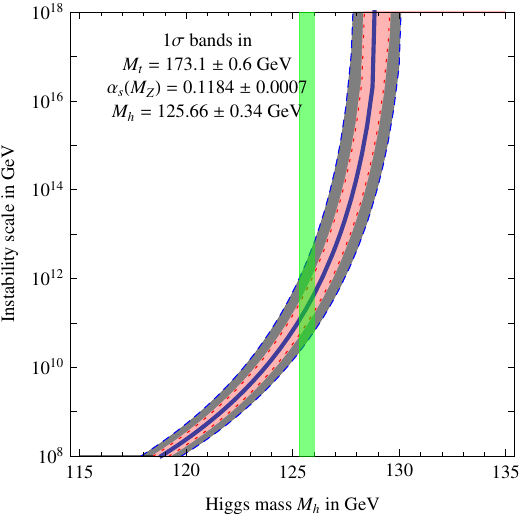}\qquad
\includegraphics[width=0.45\textwidth]{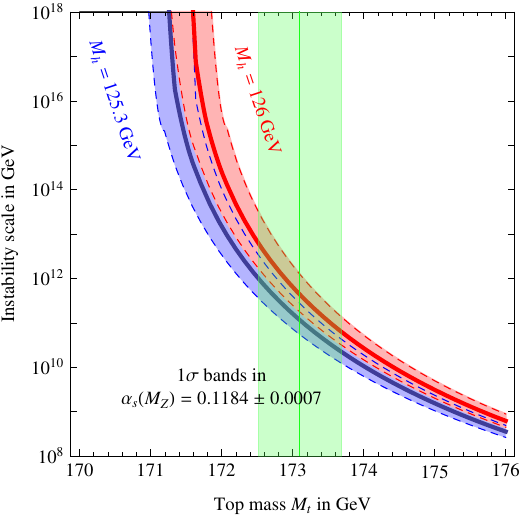}$$
\caption{\em The instability scale $\Lambda_I$ at which the SM potential becomes negative
 as a function of the Higgs mass (left) and of the top mass (right).
The theoretical error is not shown and corresponds to a $\pm1\GeV$ uncertainty in $M_h$.
\label{fig:instability}}
\end{figure}

The final result for the condition of absolute stability is presented in \eq{stability}. The central value of 
the stability bound at NNLO on $M_h$  is shifted with respect to NLO computations
(where the matching scale is fixed at $\mu=M_t$)
by 
about $+0.5\GeV$, whose main contributions can be decomposed as follows: \\[6pt]
 $~\qquad +0.6\GeV$  due to the QCD threshold corrections to $\lambda$ (in agreement with~\cite{Bezrukov:2012sa});\\ 
 $~\qquad +0.2\GeV$ due to the  Yukawa threshold corrections to $\lambda$; \\ 
 $~\qquad -0.2\GeV$ from RG equation at 3 loops (from~\cite{betagauge3,Chetyrkin:2012rz});\\ 
 $~\qquad -0.1\GeV$ from the effective potential at 2 loops.\\[6pt]
As a result of these corrections, the instability scale is lowered by a factor $\sim 2$, for $M_h\sim 125$ GeV, after including NNLO effects. The value of the instability scale is shown in \fig{fig:instability}.

\begin{figure}[t]
$$\includegraphics[height=0.32\textwidth]{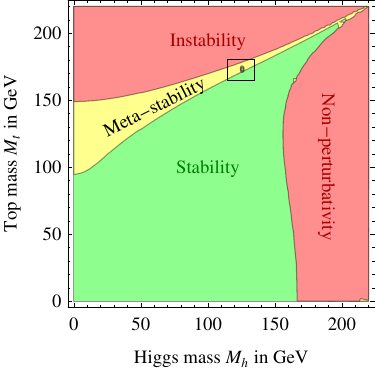}\qquad
\includegraphics[height=0.32\textwidth]{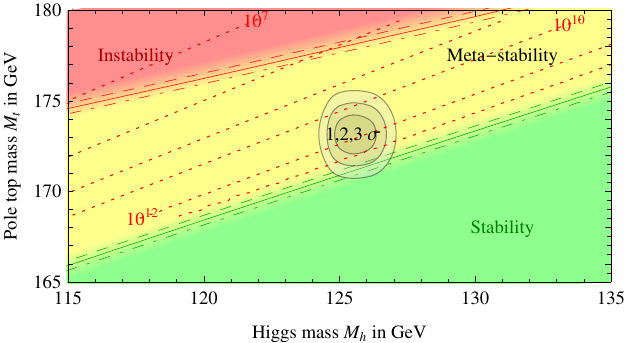}$$
\caption{\em Regions of absolute stability, meta-stability and instability of the SM vacuum
in the $M_t$--$M_h$ plane (upper left) and in the $\lambda$--$y_t$ plane, in terms of parameter renormalized
at the Planck scale (upper right).
{\bf Bottom}: Zoom in the region of the preferred experimental range of $M_h$ and $M_t$
(the gray areas denote the allowed region at 1, 2, and 3$\sigma$).
The three  boundary lines correspond to $\alpha_s(M_Z)=0.1184\pm 0.0007$, and the grading of the 
colors indicates the size of the theoretical error.  
The dotted contour-lines show the instability scale $\Lambda$ in $\GeV$ assuming $\alpha_s(M_Z)=0.1184$.
\label{fig:regions}}
\end{figure}

The phase diagram of the SM Higgs potential is shown in \fig{fig:regions}, 
taking into account the values for $M_h$ measured by 
ATLAS and CMS~\cite{ATLAS:2012ac,Chatrchyan:2012tx}. 
The left plot illustrates the remarkable coincidence for which the SM appears to live right at the border between the stability and instability regions. As can be inferred from the bottom plot, which zooms into the relevant region, there is significant preference for
meta-stability of the SM potential. By taking into account all uncertainties, we find that the stability region
is disfavored by present data by $2\sigma$. For $M_h<126$~GeV, stability up to the Planck mass is excluded at 
98\% C.L. (one sided).

\begin{table}[t]
\begin{center}
\begin{tabular}{ccc}   
Type of error &  Estimate of the error &  Impact on $M_h$  \\  \hline
$ M_t$          &  experimental uncertainty in $M_t$  &   $\pm 1.4$~GeV  \\  
$\alpha_{\rm s}$ & experimental uncertainty in $\alpha_{\rm s}$  &   $\pm 0.5$~GeV  \\ 
\color{red}\bf Experiment &  \color{red}\bf Total combined in quadrature &  \color{red}$\pm 1.5$~GeV \\   \hline 
$\lambda$   &  scale variation in $ \lambda$   &   $\pm 0.7$~GeV  \\ 
$ y_t$          &  ${\cal O}(\Lambda_{\rm QCD})$ correction to $M_t$  &   $\pm 0.6$~GeV  \\ 
$ y_t$          &  QCD threshold at 4 loops  &   $\pm 0.3$~GeV  \\ 
 RGE                    &  EW at 3 loops + QCD at 4 loops   &   $\pm 0.2$~GeV     \\   
\color{red}\bf Theory &  \color{red}\bf Total combined in quadrature & \color{red} $\pm 1.0$~GeV \\   \hline 
\end{tabular}
\caption{\label{tab:errors}\em 
Dominant sources of experimental and theoretical errors  in the computation of the SM stability bound on the Higgs mass, \eq{stability}.}
\end{center}
\end{table}

The dominant uncertainties   in the evaluation of the minimum $M_h$ value ensuring absolute
vacuum stability within the SM are summarized in 
Table~\ref{tab:errors}. 
The dominant uncertainty is experimental and comes mostly from the measurement of $M_t$.
Although experiments at the LHC are expected to improve the determination of $M_t$, the error on the top mass will remain as the largest source of uncertainty. If no new physics other than the Higgs boson is discovered at the LHC, the peculiarity of having found that the SM parameters lie at the critical border between stability and metastability regions provides a valid motivation for improved top quark mass measurements, possibly at a linear collider. 

The  dominant theoretical  uncertainty, while reduced by about a factor of 3 with the present work,
is  still related to threshold corrections to the Higgs coupling $\lambda$ at the weak scale.
Another sizable  theoretical  uncertainty comes from the fact that the pole top mass 
determined at hadron colliders suffers from ${\cal O}(\Lambda_{\rm QCD})$ non-perturbative uncertainties~\cite{Hoang:2008xm}.  
A possibility to overcome this problem and, at the same time,  to improve the experimental error on $M_t$,  
would be a direct determination of the $\overline{\rm MS}$ top-quark running mass 
from experiments, for instance from the $t \bar t$ cross-section at a future $e^+e^-$  collider operating above the $t \bar t$ threshold. In this respect,
such a collider  could become crucial for establishing the structure of the vacuum and the ultimate fate of our universe.

As far as the RG equations are concerned, the error of $\pm 0.2$~GeV is a  conservative 
estimate, based on the parametric size of the missing terms. The smallness of this error, compared 
to the uncertainty due to threshold corrections, can be understood by the smallness of all the 
couplings at high scales: four-loop terms in the RG
equations do not compete with finite tree-loop corrections close to the electroweak scale, where the  
strong and the top-quark Yukawa coupling are large.

The LHC will be able to measure the Higgs mass with an accuracy of about 100--200 MeV, which is far better than the theoretical error with which we are able to determine the condition of absolute stability.

\begin{figure}[t]
$$\includegraphics[width=0.45\textwidth]{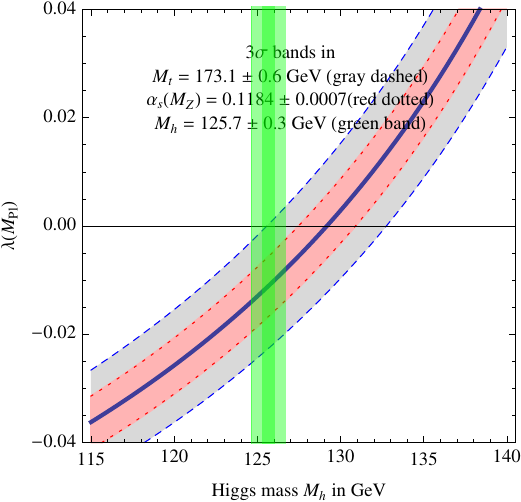}\qquad
\includegraphics[width=0.45\textwidth]{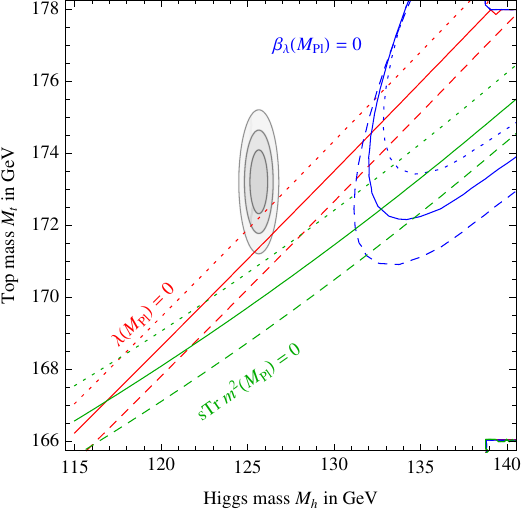}$$
\caption{\em {\bf Left}: The Higgs quartic coupling
$\lambda$ at the Planck scale as a function of $M_h$, with 
$\pm3\sigma$ variations in $M_t$ and $\alpha_{\rm s}$.
{\bf Right}: Curves in the  $M_t$--$M_h$ plane 
corresponding to the conditions $ \lambda(M_{\rm Pl})=0$ (red),
$\beta_\lambda(M_{\rm Pl})=0$ (blue), and to the 
Veltman's condition of vanishing supertrace (green, see text). In all cases the dashed and dotted lines 
denote the $\pm3\sigma$ variation in $\alpha_{\rm s}$.
The gray areas denote  the experimentally allowed region at 1, 2, and 3$\sigma$.
\label{fig:Planck}}
\end{figure}

\section{Implications}
\label{sec:num}

\subsection{Boundary conditions at the Planck scale}

It is certainly a remarkable coincidence that both $\lambda$ and its beta function $\beta_\lambda$ nearly vanish around the Planck scale. This motivates us to explore in more detail the boundary conditions at $M_{\rm Pl}$ required to reproduce the measured values of the SM parameters. In \fig{fig:Planck} (left) we show the prediction for the Higgs quartic coupling $\lambda$ at $M_{\rm Pl}$, with 3$\sigma$ bands describing the errors in $M_t$ and $\alpha_s$. For $M_h$ in the range preferred by LHC, $\lambda (M_{\rm Pl}) =0$ can be obtained only if $M_t$ is 
$\approx 2$~GeV below its present central value (\ie\ $M_t\le 171$~GeV). However, it should be noted that $\lambda =0$ is neither a fixed point nor a point of enhanced symmetry and thus satisfying {\it exactly} this condition is not especially meaningful. The best-fit value for $\lambda$ at the Planck scale is small and negative, 
\beq \lambda (M_{\rm Pl})=-0.0129+ 0.0028 \left(\frac{M_h}{\GeV}-125.5\right)
\pm 0.0047_{M_t}\pm 0.0018_{\alpha_s}\pm 0.0028_{\rm th}\label{lambdaPl}\ ,\eeq 
where the uncertainties refer to
the present $1\sigma$ errors in $M_t$ and $\alpha_{\rm s}$, and to the theoretical error.

Equation~(\ref{lambdaPl}) could be the result of matching the SM at $M_{\rm Pl}$ with a theory in which $\lambda$ vanishes at tree level and receives one-loop threshold corrections. However,
the contribution to high-energy thresholds from the SM couplings at $M_{\rm Pl}$ is typically a few times $10^{-3}$. 
The required effect of size $10^{-2}$ can be obtained from loops of SM couplings only if these involve particles with large multiplicities, or else from loop with new (large) coupling constants. As evident from \fig{fig:run1}, the boundary condition of $\lambda$ at high energy is fairly independent of the precise value at which we impose it. 

The right plot in \fig{fig:Planck} shows the dependence on $M_h$ and $M_t$ of the boundary conditions $\lambda (M_{\rm Pl})=0$ (red line) and $\beta_\lambda (M_{\rm Pl})=0$ (blue line). While $\lambda =0$ weakly depends on the scale at which is evaluated, a more pronounced dependence affects the condition $\beta_\lambda =0$ (see \fig{fig:run2}).  This is because $\beta_\lambda$ depends not only on $\lambda$, but also on other couplings (top Yukawa and gauge) that run in the high-energy region. As a result, although $\beta_\lambda (M_{\rm Pl})=0$ cannot be exactly satisfied, the  beta function vanishes at scales very near the Planck mass. For instance, for $M_t=171.0$~GeV and $M_h=125$~GeV, both $\lambda$ and $\beta_\lambda$ simultaneously vanish when they are evaluated at a scale equal to $3\times10^{17}\GeV$.
In ref.~\cite{Shaposhnikov:2009pv} it was argued that $\lambda (M_{\rm Pl})\approx 0$ and $\beta_\lambda (M_{\rm Pl})\approx 0$ could be justified in the case of an asymptotically safe gravitational theory.

Just for illustration,  in the right plot in \fig{fig:Planck} we also show the Higgs mass implied by
Veltman's condition~\cite{Veltman:1980mj} that the supertrace of the squared masses of all SM particles vanishes at a given scale, here chosen to be $M_{\rm Pl}$: ${\rm STr} {\cal M}^2 (M_{\rm Pl})=0$. 
We remark, however, that this condition does not carry special information about the power divergences of the theory, which are dominated by UV effects, while the supertrace includes only the contribution from the IR degrees of freedom in the SM. 
At any rate, the possibility 
of a very special fine-tuning involving only the SM loop contributions and leading to
${\rm STr} {\cal M}^2 (M_{\rm Pl})=0$, implies a Higgs mass $M_h\approx (135\pm 2.5)\GeV$,
which is excluded at more than 3$\sigma$.
Lowering the scale at which the supertrace condition is evaluated makes the disagreement even stronger.

\subsection{Higgs inflation from non-minimal coupling to gravity}
The extrapolation of the SM up to very high energy has led to some speculations about the possibility of interpreting the Higgs boson as the inflaton. 
One scenario for Higgs inflation~\cite{Bezrukov:2007ep} exploits a large non-minimal coupling between the Higgs bilinear and the Ricci scalar $R$, with an interaction Lagrangian $\xi |H|^2 R$. The effect of this interaction is to flatten the Higgs potential 
(or any other potential) above the scale $M_{\rm Pl}/\sqrt{\xi}$, providing a platform for slow-roll inflation. A correct normalization of the spectrum of primordial fluctuations fixes the value of the coupling constant $\xi$. Using the tree-level potential, one finds $\xi \approx  5\times 10^4\sqrt{\lambda}$.

This inflationary scenario, attractive for its minimality, suffers from a serious drawback. Perturbative unitarity is violated at the scale $M_{\rm Pl}/\xi$, signaling the presence of new physical phenomena associated with strong dynamics. It is naturally expected that these phenomena will affect the scalar potential above $M_{\rm Pl}/\xi$ in an uncontrollable way~\cite{Burgess:2009ea}. One solution is to add new degrees of freedom that restore perturbative unitarity~\cite{Lerner:2010mq}, although the minimality of the model is then lost. The procedure advocated by the proponents of this scenario is to assume that the strong dynamics will preserve intact the shape of the SM potential, even above $M_{\rm Pl}/\xi$. Although we find this assumption questionable, it is still interesting to address the issue of whether the Higgs data are compatible with this scenario.

A two-loop analysis of Higgs $\xi$-inflation was developed in ref.~\cite{Bezrukov:2009db,DeSimone:2008ei}. The ordinary SM evolution is perfectly adequate below the scale $M_{\rm Pl}/\xi$, while the new interaction can affect the scalar potential at very high energy. The renormalization procedure above the inflationary scale is not unambiguous as, for instance, the renormalization scale differs in the Jordan and Einstein frames. Luckily, the slow running of $\lambda$ at high energy makes these issues irrelevant, from a practical point of view. A simple SM calculation of the potential is perfectly adequate to describe the situation of Higgs $\xi$-inflation (see also the discussion in ref.~\cite{Bezrukov:2012sa}).

In practice, the result is that Higgs $\xi$-inflation requires stability of the potential up to the inflationary scale $M_{\rm Pl}/\sqrt{\xi}$. As we are interested in the minimum value of the Higgs mass that satisfies this condition, the coupling $\lambda$ at the relevant scale is very small and thus the coupling $\xi$ is not particularly large, $\xi < {\cal O}(10^3)$. 
Therefore, the resulting restriction is stability, as given by \eq{stability}. If the LHC indication for $M_h=125$--126 GeV is confirmed, the simplest version of Higgs inflation is disfavored, unless the top mass is about 2$\sigma$ below its present central value. However, given the proximity of $\lambda(M_{\rm Pl})$ to the critical value for stability, unknown
one-loop threshold corrections near the Planck mass could be sufficient to rescue the proposal. It is also interesting that the introduction of a single scalar field at the scale $M_{\rm Pl}/\xi$ could simultaneously restore perturbative unitarity and cure the potential instability~\cite{EliasMiro:2012ay}.

\subsection{Higgs inflation from false vacuum}
Alternative proposals for Higgs inflation employ the peculiarity of the SM scalar potential to develop a second minimum at large Higgs field values for a very special choice of parameters \cite{Nielsen,BdCE}. 
The possibility of using this new minimum for inflation was first contemplated in ref.~\cite{Isidori:2007vm}, finding that it
implies a viable prediction for the Higgs mass, but also a wrong prediction for the amplitude of density fluctuations.
The latter result can be cured in non-minimal inflationary setups~\cite{Masina:2011aa,Masina:2012yd}
without affecting the prediction for the Higgs mass, which we now precisely compute.

The first derivative of the Higgs potential $V=\lambda_{\rm eff} (h) h^4/4$ is
\be
\frac{dV}{dh} =\left( \lambda_{\rm eff} + \frac{\beta_{\rm eff}}{4}\right) h^3~.
\ee
Here $\lambda_{\rm eff} (h)$ is the effective coupling defined in \eq{lambdaeff} and $\beta_{\rm eff} = d \lambda_{\rm eff}/d \ln h$. If $\lambda_{\rm eff}$ becomes sufficiently small, the potential can develop a minimum at $h=h_{\rm min}$, such that
\be
\left. \lambda_{\rm eff} +\frac{\beta_{\rm eff}}{4}\right|_{h=h_{\rm min}}=0~.
\ee
This situation can occur in the proximity of a field value $h_*$ where $\beta_{\rm eff}$ vanishes. In the neighborhood of $h_*$, we can approximate $\lambda_{\rm eff} (h)$ as
\be
\lambda_{\rm eff} (h) \approx \lambda_* + b \ln^2 \frac{h}{h_*}~,
\label{lamapp}
\ee
where $\lambda_*$ is the minimum value of $\lambda_{\rm eff}$, such that $\beta_{\rm eff}(\lambda_*)=0$. The zero of the $\beta$ function insures that the leading log is absent in \eq{lamapp} and thus $b$ is a typical two-loop coefficient. For the relevant values of the SM parameters, we find
 $b=0.4/ (4 \pi)^{4}$. We are interested in a situation in which the field configuration corresponds to a local minimum (while the EW vacuum remains the global minimum) and thus we want both $\lambda_*$ and $b$ to be positive. Using the expansion in \eq{lamapp}, we can compute $h_{\rm min}$ and the minimum of the potential $V_{\rm min} \equiv V(h_{\rm min})$,
\be
h_{\rm min} \approx h_* \exp \left[ \frac 14 \left( \sqrt{1-\frac{16 \lambda_*}{b}}-1\right) \right] ~,
\ee
\be
V_{\rm min} \approx \frac{b}{8}~h_{\rm min}^4 \ln \frac{h_*}{h_{\rm min}}~.
\ee
The minimum $h_{\rm min}$ exists only for extremely small values of the Higgs quartic coupling, $\lambda_*<b/16$. As we vary $\lambda_*$ within its allowed range ($0<\lambda_*<b/16$) we find that $h_{\rm min}$ is always near $h_*$ ($e^{-1/4}<h_{\rm min}/h_*<1$), while $V_{\rm min}$ can change widely ($0<V_{\rm min}<bh_{\rm min}^4/32$).

If the Higgs field is trapped in the false vacuum during the early universe, it can cause inflation. The normalization of the spectrum of primordial perturbations, which is determined by $V_{\rm min}$, can be appropriately selected by tuning the ratio $\lambda_*/b$. The main difficulty of this scenario is to achieve a graceful exit from the inflationary phase. Two mechanisms have been proposed. The first one~\cite{Masina:2011aa} employs a new scalar field, non-minimally coupled to gravity, that slows down the expansion rate, thus allowing for quantum tunneling of the Higgs out of the false vacuum. The second mechanism~\cite{Masina:2012yd} uses a scalar field weakly coupled to the Higgs which, during the cosmological evolution, removes the barrier in the Higgs potential in a process analogous to hybrid inflation. So, in practice, the minimality of the SM is lost and one may wonder if there is any conceptual gain with respect to adding a new scalar playing the role of the inflaton. Nevertheless, it is interesting to investigate whether the Higgs and top masses are compatible with the intriguing possibility of a false vacuum at large field value.

\begin{figure}[t]
$$\includegraphics[width=\textwidth]{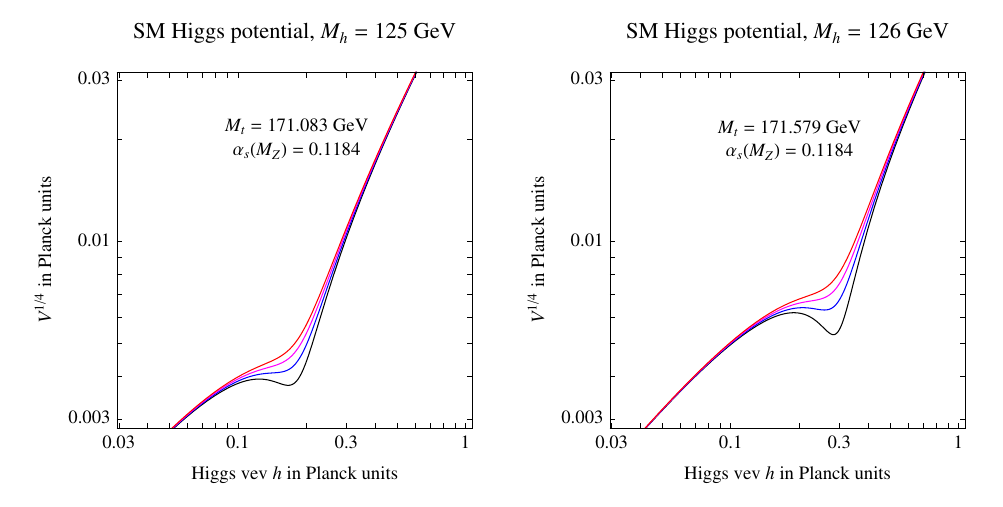}$$
\caption{\em Two-loop SM Higgs potential for $M_h=125,126\GeV$ around the critical top mass
that gives a second minimum around the Planck scale.
The various curves correspond to variations in $M_t$ by $0.1\MeV$.
\label{fig:V}}
\end{figure}

In practice the above equations amount to saying that
the conditions for the existence of a second (unstable) vacuum are that $\lambda_{\rm eff}$ (essentially) vanishes at the same scale at which $\beta_{\rm eff}$ vanishes. 
This corresponds to the intersection between the red band (condition $\lambda \approx \lambda_{\rm eff}=0)$ and the
blue band (condition $\beta_\lambda=0)$ in \fig{fig:Planck}(right).
It is remarkable that the SM can achieve these conditions, although they require a top mass about 2$\sigma$ below the central value. 
The resulting relation between $M_h$ and $M_t$ corresponds to the equality in \eq{stability}, and
is precisely studied in \fig{fig:V} 
where we compute  for $M_h=\{125,126\}\GeV$ the predicted top mass 
and show the shape of the potential around the false vacuum. 
The value of $V_{\rm min}$ can be changed by tuning $\lambda_*$ or, in other words, by accurate variations of $M_h$ and $M_t$. The existence of the false vacuum depends critically on the exact values of the SM parameters and requires dialing $M_h$ and $M_t$ by one part in $10^6$. However, the exact value of the needed top mass has a theoretical uncertainty, reduced down to $\pm0.5\GeV$
thanks to our higher-order computation.
Note from \fig{fig:V} that the field value where the false vacuum is positioned is larger than what was reported in~\cite{Isidori:2007vm,Masina:2011aa}. The corrections in \eq{lambdaeff} \cite{CEQ,BdCE} are mostly responsible for the larger field values found in our analysis.

\begin{figure}[t]
$$
\includegraphics[width=0.7\textwidth]{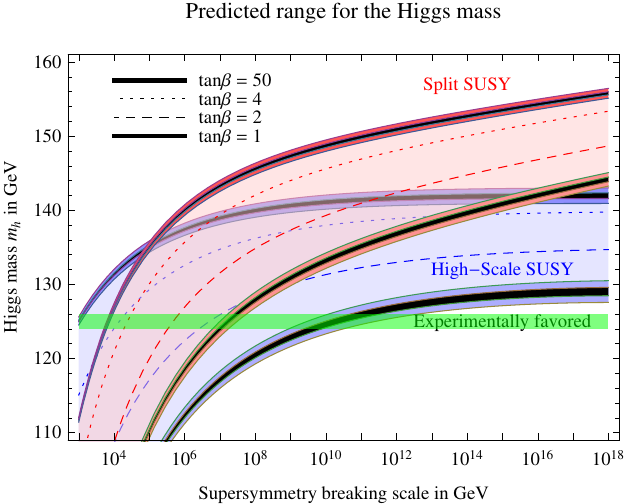}$$
\caption{\em NNLO prediction for the Higgs mass $M_h$ in High-Scale Supersymmetry (blue, lower) and Split Supersymmetry (red, upper) 
for $\tan\beta=\{1,2,4,50\}$. The thickness of the lower boundary at $\tan\beta=1$ and of the upper boundary at $\tan\beta=50$ 
shows the uncertainty due to the present $1\sigma$ error on $\alpha_s$ (black band) and on the top mass (larger colored band).
\label{fig:mhrange}}
\end{figure}

\subsection{Supersymmetry}
Our higher order computation of the relation between the Higgs mass and the Higgs quartic coupling $\lambda$ 
has implications for any model that can predict $\lambda$.
If supersymmetry is present at some scale $\tilde m$, then in the minimal model one finds the tree-level relation
   \beq
\lambda (\tm )= \frac18\left[g^2(\tm)+g^{\prime 2}(\tm)\right] \cos^22\beta ~.
\label{match2}
\eeq 
A dedicated analysis of the resulting prediction for the Higgs mass as function of $\tilde m$ and of $\tan\beta$ was performed
in~\cite{Giudice:2011cg} (see also \cite{CCD}).  We here update the results, including the new correction which increases the predicted Higgs mass by an amount that changes with $M_h$. 
Once more, the main impact of our calculation is the reduction of the theoretical uncertainty from $\pm 3\GeV$ down to $\pm 1\GeV$.
As a consequence, supersymmetry broken at the Planck scale, which requires
$\lambda\ge0$ and thus the stability condition \eq{stability}, is disfavored at
$2\sigma$, unless thresholds at $M_{\rm Pl}$ (or non-minimal couplings) account for the small negative value of $\lambda(M_{\rm Pl})$, see \eq{lambdaPl}.
Thresholds at one loop have been computed in~\cite{Giudice:2011cg} including finite parts and the constant
factor due to the fact that the $\overline{\rm MS}$ renormalization scheme breaks supersymmetry.
A simplified, but illustrative, formula for the supersymmetric threshold corrections is obtained by taking a common mass $M_2$ for weak gauginos and higgsinos, and a common mass $m_{\tilde{t}}$ for the stops,
\beq 
\delta \lambda(M_{\rm Pl}) \approx \frac{1}{(4\pi)^2}\left[ -0.25 + 0.12 \ln\frac{m_{\tilde{t}}}{M_2} + 0.05\ln\frac{m_A}{M_2} \right] ~,
\label{thzuzy}
\eeq
where $m_A$ is the mass of the Higgs pseudoscalar. The absence of a scale dependence in \eq{thzuzy} is a consequence of the approximate cancellation of $\beta_\lambda$ around $M_{\rm Pl}$. Negative values of the boundary condition for $\lambda$ thus require stops lighter than higgsinos, winos, and bino. By using the full formula and allowing for each sparticle mass to vary by one order of magnitude above or below the average mass $\tm$ we find
\beq
-0.006< \lambda(M_{\rm Pl}) <0.002~.
\eeq
This is insufficient to reach the central value of \eq{lambdaPl} and thus indicates that supersymmetry has to be broken at a scale below $M_{\rm Pl}$, if the Higgs mass has to match a supersymmetric boundary condition.

Our predictions for the Higgs mass as a function of the supersymmetry breaking scale $\tilde m$ are illustrated in \fig{fig:mhrange}, in the case of High-Scale Supersymmetry (all supersymmetric particles with masses equal to $\tilde m$) and Split Supersymmetry~\cite{split} (supersymmetric scalars with masses equal to $\tilde m$ and supersymmetric fermions at the weak scale). We refer the reader to ref.~\cite{Giudice:2011cg} for details about the underlying assumptions of the calculation. 

Figure \ref{fig:mhrange} shows not only how $M_h\approx 125\GeV$ disfavors supersymmetry broken at a very high scale, but also the well know fact that the usual
scenario of weak-scale supersymmetry can account for the Higgs mass
only for extreme values of the parameters (such as large $\tan\beta$, heavy stops, maximal
stop mixing). In the case of Split Supersymmetry, large values of $\tilde m$ are clearly excluded by LHC Higgs searches. On the other hand, (mildly ``unnatural") scenarios in which the masses of supersymmetric scalars are one-loop larger than the weak scale~\cite{anmod} are in good agreement with data.

\section{Conclusions}

One of the most important questions addressed by the LHC is
naturalness. Not only will the answer affect our understanding of the
mechanism for EW breaking, but it will also determine our strategy for
future directions in theoretical physics. On one side we have the
avenue of larger symmetries unifying physical laws in a single
fundamental principle; on the other side we have new kinds of
paradigms, where parameters are not understood by naturalness
arguments in the context of well-defined effective theories. At
present, the multiverse is one of the most intriguing options to
pursue the latter path.

If the LHC finds Higgs couplings deviating from the SM prediction and
new degrees of freedom at the TeV scale, then the most important
question will be to see if a consistent and natural (in the technical
sense) explanation of EW breaking emerges from experimental data. But
if the LHC discovers that the Higgs boson is not accompanied by any
new physics, then it will be much harder for theorists to unveil the
underlying organizing principles of nature. The multiverse, although
being a stimulating physical concept, is discouragingly difficult to
test from an empirical point of view. The measurement of the Higgs
mass may provide a precious handle to gather some indirect
information.

Once we extrapolate the SM to very short distances, we find that the
values of the Higgs mass, hinted by the first LHC results (125--126
GeV), lie right at the edge between EW stability and instability
regions, see \fig{fig:regions}. Moreover, the slow running of the
quartic Higgs coupling $\lambda$ in the high-energy regime implies
that the instability scale most critically depends on the Higgs mass
$M_h$. A small change in $M_h$ (and $M_t$) can drastically modify our
conclusions regarding vacuum stability. This special situation
motivated us to perform a  NNLO calculation of the Higgs
potential in the SM, which is the subject of this paper.

Our calculation includes three-loop running for gauge, top Yukawa, and Higgs quartic coupling and two-loop matching conditions keeping the leading effects in $\alpha_s$, $y_t$ and $\lambda$. In particular, we have computed for the first time the two-loop threshold correction to $\lambda$, which was the most sizable missing ingredient of the NNLO result. The completion of the NNLO calculation allows us to reduce the theoretical error in the stability limit on the Higgs mass from 3~GeV to 1~GeV. Our final result is shown in \eq{stability}. After our calculation, the largest source of uncertainty comes from the parametric dependence on the top quark mass, which leads to a 1$\sigma$ error of 1.4~GeV in the critical Higgs mass. Of course our calculation reliably accounts for IR effects, but ignores possible new (unknown) UV threshold effects occurring near the Planck scale. Since our goal is to learn information about physics at very short distances, the high-energy corrections constitute an opportunity, rather than a deficiency in the calculation. 

The first lesson that we learn from the SM extrapolation is that the Higgs mass hinted by LHC results corresponds to $\lambda \approx 0$ and $\beta_\lambda \approx 0$ at high energies. This, by itself, is an intriguing result because $\lambda =0$ is the critical value for stability and it may hide some information about Planckian physics. With our precise calculation, we can investigate further the situation. We find that, for $M_h =125$~GeV, $\lambda (M_{\rm Pl})=-0.014\pm 0.006$, see \eq{lambdaPl}. The exact vanishing of $\lambda (M_{\rm Pl})$ is excluded at 2$\sigma$. Moreover, the smallness of $\beta_\lambda$ at high energy implies that $\lambda$ remains negative in a relatively large energy range. Indeed, we find that, for $M_h =125$~GeV, the instability scale develops at $10^{11\pm1}$~GeV. Quantum tunneling is sufficiently slow to ensure at least metastability of the EW vacuum, see \fig{fig:regions}. The small value of $\lambda (M_{\rm Pl})$ may indicate a radiative origin, although typical one-loop effects of SM couplings appears to be insufficient to account for it. We have also shown that $\beta_\lambda$ varies more rapidly at high energy and vanishes at a scale of about $3\times 10^{17}\GeV$.

The stability of the SM potential is a crucial issue for models of inflation that employ the Higgs boson. We have analyzed several proposals showing that present data disfavor them at 98\% C.L. These models can still be viable if the top quark mass turns out to be less than about 171~GeV or if new physics around $M_{\rm Pl}$ slightly modifies the shape of the Higgs potential. The latter possibility, although fairly plausible, limits the predictability and the minimality of the approach. We have also updated previous predictions for the Higgs mass in High-Scale Supersymmetry and Split Supersymmetry. 

It is natural to try to speculate on the possible meaning of the near vanishing of $\lambda$ and $\beta_\lambda$ around the Planck scale. The coupling $\lambda =0$ is the critical value that separates the ordinary EW phase from a phase in which the Higgs field slides to very large values. It is noteworthy that the hierarchy problem can also be interpreted as a sign of near criticality between two phases~\cite{Giudice:2006sn}. The coefficient $m^2$ of the Higgs bilinear in the scalar potential is the order parameter that describes the transition between the symmetric phase ($m^2>0$) and the broken phase ($m^2<0$). In principle, $m^2$ could take any value between $-M_{\rm Pl}^2$ and $+M_{\rm Pl}^2$, but quantum corrections push $m^2$ away from zero towards one of the two end points of the allowed range. The hierarchy problem is the observation that in our universe the value of $m^2$ is approximately zero or, in other words, sits near the boundary between the symmetric and broken phases. Therefore, if the LHC result is confirmed, we must conclude that both $m^2$ and $\lambda$, the two parameters of the Higgs potential, happen to be near critical lines that separate the EW phase from a different (and inhospitable) phase of the SM. We do not know if this peculiar quasi-criticality of the Higgs parameters is just a capricious numerical coincidence or the herald of some hidden truth.

The occurrence of criticality could be the consequence of symmetry. For instance, supersymmetry implies $m^2=0$. If supersymmetry is marginally broken, $m^2$ would remain near zero, solving the hierarchy problem. But if no new physics is discovered at the LHC, we should turn away from symmetry and look elsewhere for an explanation of the near-criticality of $m^2$.

The critical value $\lambda =0$ could be justified by symmetry reasons. For instance, if the Higgs is a Goldstone boson, its potential vanishes and both $m^2$ and $\lambda$ are zero. The non-vanishing top Yukawa coupling prevents this possibility to be realized exactly. Radiative corrections then completely spoil the solution to the hierarchy problem, but could generate a small and negative value of $\lambda$ at the Planck scale, compatible with our results. Supersymmetry broken at high scales could also account for a vanishing boundary condition of $\lambda$, if $\tan\beta =1$. However, we have shown that, unless the Higgs has strong couplings with new states that live at Planckian energy, the supersymmetric boundary condition cannot be satisfied at $M_{\rm Pl}$, see \fig{fig:mhrange}. 

Alternatively, criticality could be the consequence of dynamics. If transplanckian dynamics induce a large anomalous dimension for the Higgs field, the matching condition at $M_{\rm Pl}$ of the quartic coupling $\lambda$ could be very small, while the top Yukawa coupling remains sizable. It was argued in ref.~\cite{Shaposhnikov:2009pv} that gravity itself could be responsible for a large anomalous dimension of the Higgs in the transplanckian region.

It is known that statistical systems often approach critical behaviors as a consequence of some internal dynamics or are attracted to the critical point by the phenomenon of self-organized criticality~\cite{SOC}. As long as no new physics is discovered, the lack of evidence for a symmetry explanation of the hierarchy problem will stimulate the search for alternative solutions. The observation that both parameters in the Higgs potential are quasi-critical may be viewed as evidence for an underlying statistical system that approaches criticality. The multiverse is the most natural candidate to play the role of the underlying statistical system for SM parameters. If this vision is correct, it will lead to a new interpretation of our status in the multiverse: our universe is not a special element of the multiverse where the parameters have the peculiarity of allowing for life, but rather our universe is one of the most common products of the multiverse because it lies near an attractor critical point. In other words, the parameter distribution in the multiverse, instead of being flat or described by simple power laws (as usually assumed) could be highly peaked around critical lines because of some internal dynamics. Rather than being selected by anthropic reasons, our universe is simply a very generic specimen in the multitude of the multiverse.

The indication for a Higgs mass in the range 125--126~GeV is the most
important result from the LHC so far. If no new physics at the TeV
scale is discovered, it will remain as one of the few and precious
handles for us to understand the governing principles of nature. The
apparent near criticality of the Higgs parameters may then contain
information about physics at the deepest level.

\subsubsection*{Acknowledgments}
J.R.E. thanks D.R.T. Jones and M. Steinhauser for useful information
exchanges. J.R.E and J.E.-M. thank CERN for hospitality and partial
financial support.  G.D. thanks P. Slavich for useful discussions.
This work was supported by the ESF grant MTT8; by
SF0690030s09 project; by the EU ITN �Unification in the LHC Era�,
contract PITN- GA-2009-237920 (UNILHC); the Spanish Ministry MICINN
under contracts FPA2010-17747 and FPA2008-01430; the Spanish
Consolider-Ingenio 2010 Programme CPAN (CSD2007-00042); and the
Generalitat de Catalunya grant 2009SGR894; by the Research Executive 
Agency (REA) of the European Union
under the Grant Agreement number PITN-GA-2010-264564 (LHCPhenoNet).

\appendix

\section*{Appendix: SM Effective Potential up to two-loops}
\label{app:potential}

The SM effective potential is known up to two-loops \cite{V2}.
We  present here its explicit expression in the $\overline{\rm MS}$ scheme and the Landau gauge.
The tree-level part is (in this section we denote the Higgs field by $\phi$ to avoid confusion with the symbol $h$ defined later)
\be
V_0(\phi) = -\frac{1}{2}m^2\phi^2 +\frac{1}{4}\lambda\phi^4\ .
\ee

The one-loop Coleman-Weinberg potential \cite{CW} is
\bea
V_1(\phi)&=&\frac{\kappa}{4}\left[-12 m_t^4(L_t-3/2)+6m_W^4(L_W-5/6)+3m_Z^4(L_z-5/6)\right.
\nonumber\\
&&\left.+m_h^4(L_h-3/2)+3m_\chi^4(L_\chi-3/2)\right]
\label{CW}
\eea
where $\kappa=1/(16\pi^2)$, $m_t^2=y_t^2\phi^2/2$ is the top mass squared, $m_h^2=-m^2+3\lambda \phi^2$ the Higgs mass squared, $m_\chi^2(\phi)=-m^2+\lambda \phi^2$ is the Goldstone mass squared and $L_t=\ln (m_t^2/Q^2)$, etc.

We split the two-loop potential in different pieces according to their diagrammatic origin. We use the short-hand notation $t\equiv m_t^2$, $w\equiv m_w^2$, $z\equiv m_z^2$, $h\equiv m_h^2$, $\chi\equiv m_\chi^2$ and we neglect the bottom Yukawa coupling.
The important top Yukawa contribution is
\bea
V_Y&=&
\frac{3}{2} y_t^2 \kappa ^2 \left[2 J_{tt}-4 J_{t\chi}-2 J_{th}+(4t-h) I_{tth}+2(t-\chi ) I_{t\chi 0}-\chi I_{tt\chi}  \right]
\ .\label{VY}
\eea
There is a purely scalar piece
\be
V_S=\frac{3}{4} \kappa ^2 \lambda  \left[5 J_{\chi\chi}+2 J_{h\chi}+J_{hh}-4 \lambda \phi^2 (I_{h\chi\chi}+I_{hhh}) \right]\ ,
\ee
a purely gauge part
\bea
V_V&=& \frac{e^2}{4z} \kappa ^2(z-w)\left[J_{zw}+w(I_{zw0} -I_{w00})\right]
\nonumber\\
&-&\frac{e^2}{4} \kappa ^2w\left[2(11 A_{z}-25A_{w})+\frac{1}{w}(24 J_{ww}+25 J_{zw})+24 I_{zww}+10I_{zw0} -9I_{w00}+49 w\right]
\nonumber\\
&+&\frac{g^2}{4} \kappa ^2w\left[\frac{58}{3} (A_{z}+2A_{w})+\frac{1}{w}(7J_{ww}+ 15 J_{zw}) +58 I_{zww}-9 I_{zw0}+I_{z00}+I_{w00}+76 w\right]
\nonumber\\
&+&\frac{G^2}{8} \kappa ^2\left[ J_{ww} -(16w +z) I_{zww}+2(8w+z)I_{zw0}-z I_{z00}+4 w^2 \right]
 \ ,\label{VV}
\eea
a fermion-gauge boson part\footnote{Notice that the first paper in ref.~\cite{V2} contains a typo for this piece, with an extra factor 3 for the lepton-lepton-$Z$ contributions.}  (which includes the important QCD piece)
\bea
V_{FV}&=&8g_s^2\kappa ^2  m_t^4\left(3L_t^2-8 L_t+9\right)+\frac{16}{3} e^2\kappa ^2\left(t A_z+J_{tz}-t I_{tt0}+t  I_{ttz} \right)
\nonumber\\
&+&\frac{g^2}{6} \kappa ^2\left\{9 t^2-16 t w-36 w^2-26t A_{t}+6(4w-3 t) A_{w}+8( t+4 w)A_{z}-4 J_{tt}+8 J_{tz}
\right.\nonumber\\
&+&\left.8(t-2 w) I_{ttz}-54 w I_{w00}-80w I_{z00}
+\frac{9}{w} \left[(t-2w) J_{tw}+(t-w)(t+2 w) I_{tw0}- t^2I_{t00}\right]\right\}\nonumber\\
&+&\frac{G^2}{6} \kappa ^2\left\{-t A_t-(17 t+40 w-20 z) A_z+\frac{17}{2} J_{tt}-17 J_{tz}
-\frac{1}{2}(7 t-40 w+17 z) I_{ttz}\right.\nonumber\\
&+&\left.\left(100w -\frac{103}{2} z\right)I_{z00}+9 t^2+20 t w-48 w^2-4 t z+60 w z-30 z^2\right\}
\ ,\label{VFV}
\eea
and a scalar-gauge boson part
\bea
V_{SV}&=&g^2 \kappa ^2\left\{\left[
\frac{1}{2} (h+3 \chi +z)-\frac{1}{3} w \right] A_{w}+\frac{3}{2} w (A_{h}+A_{\chi})
+\frac{1}{4} (J_{\chi\chi}+J_{h\chi})+
\frac{ (h-w)^2}{4 w}I_{wh0}\right.\nonumber\\
&+&
\frac{1}{4 w}\left[\frac{1}{2}(h-2 w) J_{ww}+
(3 w+\chi-h )J_{wh}+ (h+5 w+z-\chi )J_{w\chi}\right]-\left(\frac{w}{4}- \chi \right)I_{w\chi\chi}\nonumber\\
&-&\left.\frac{1}{8 w} \left(h^2-4 h w+12 w^2\right) I_{wwh}- \left[\frac{1}{4w}(h+w-\chi )^2-h \right]I_{wh\chi}-w\left(w+\frac{h}{2}\right)\right\}\nonumber\\
&+&\frac{1}{2}\left\{\begin{array}{c}
w\leftrightarrow z\\
g\leftrightarrow G
\end{array} \right\}+\frac{3g^2}{16w} \kappa ^2\left(8\lambda ^2\phi ^4I_{h\chi 0}-h^2 I_{h00}\right)\nonumber\\
&-&
\frac{e^2}{4 w z} (z-w)\kappa ^2\left\{(w+z-\chi )J_{zw}-w J_{z\chi}+ \left[(w+z-\chi )^2+8 w z\right]I_{zw\chi}\right.\nonumber\\
&-&\left.(w-\chi )^2I_{w\chi 0}-(z-\chi )^2I_{z\chi 0}+\chi ^2I_{\chi 00}\right\}\nonumber\\
&-&\frac{e^2}{2} \kappa ^2 \left\{
\left(4\chi +w-\frac{5}{3}z-\frac{z^2}{2w}\right)A_z+w A_w-J_{w\chi}+\left( 4-\frac{z}{4w}\right)J_{z\chi}
+\frac{3}{2}\chi ^2+2 z(z+\chi )
\right.\nonumber\\
&
-&\left.\frac{w}{2}(w+2\chi )-\frac{1}{4}\chi ^2\left(6+\pi ^2\right)+(4\chi  -z)I_{z\chi\chi}
-\frac{13}{4}\chi  I_{\chi\chi 0}+\frac{3}{2}(3 w-\chi  )I_{w\chi 0}\right\}
\ .
\eea
The functions $A$, $J$ and $I$ are
\bea
A_x\equiv A[x]&\equiv& x(L_x-1)\ ,\\
J_{xy}\equiv J[x,y]&\equiv & A[x]A[y]\ ,\\
I_{xyz}\equiv I[x,y,z]&\equiv &
\frac{1}{2}\left[(x-y-z)L_yL_z+(-x+y-z)L_xL_z+(-x-y+z)L_xL_y\right]\nonumber\\
&+&2(x L_x+y L_y+z L_z)-\frac{5}{2}(x+y+z)-\frac{1}{2}\xi[x,y,z]\ ,
\eea
where $L_x=\ln (x/Q^2)$ and
\bea
\xi[x,y,z]&=&R\left[2\ln\left(\frac{x-y+z-R}{2z}\right)\ln\left(\frac{-x+y+z-R}{2z}\right)-\ln\left(\frac{x}{z}\right)\ln\left(\frac{y}{z}\right)\right.\nonumber\\
&-&2\left.{\rm Li}_2\left(\frac{x-y+z-R}{2z}\right)-2{\rm Li}_2\left(\frac{-x+y+z-R}{2z}\right)+\frac{\pi^2}{3}\right]\ ,
\eea
where $R^2=x^2+y^2+z^2-2xy-2xz-2yz$ and ${\rm Li}_2(x)$ is the dilogarithm function. 
The above expression is valid for $R^2>0$, while for $R^2<0$ the analytical continuation should be used instead.
Some particular cases of the previous functions which are useful to evaluate 
the potential are
\bea
I[x,y,0]&=&(x-y)\left[\mathrm{Li}_2\left(\frac{y}{x}\right)-\frac{\pi^2}{6}-(L_x-L_y)L_{x-y}+\frac{1}{2}L_x^2\right]\nonumber\\
&&-\frac{5}{2}(x+y)+2x L_x+2y L_y -x L_x L_y\ ,\nonumber\\
I[x,x,0]&=&x\left(-L_x^2+4L_x-5\right)\ ,\nonumber\\
I[x,0,0]&=&x\left(-\frac{1}{2}L_x^2+2L_x-\frac{5}{2}-\frac{\pi^2}{6}\right)\ .
\eea

The 2-loop corrections to $\lambda_{\rm eff}$ including subleading corrections not present in \eq{lambdaeff} (but setting 
$\lambda\rightarrow 0$, as this coupling is particularly small at high field values) are explicitly given by
\bea
\delta \lambda_{\rm eff}&=&
\kappa ^2\left\{8g_s^2 y_t^2 (3r_t^2-8r_t+9)-
\frac{1}{6}g^2 y_t^4(12r_t^2-23r_t-9r_w+45)\right.\nonumber\\
&+&
\frac{g^6}{48} \left[-30 r_w^2-18 r_{t/w}r_{(t-w)^2/(tw)}+532 r_w+144 r_{z/w}-598+12 \pi ^2\right]
\nonumber\\
&+&
\frac{g^4 G^2}{96} \left[397-32 r_{t/z}^2+126r_{z/w}^2+66 r_z^2 
+27 r_w^2 -232 r_z-138 r_w+160 \frac{\pi ^2}{3}\right]
\nonumber\\
& +&
\frac{g^4 y_t^2}{24} \left[-27 r_w^2+27 r_{t/w}r_{(t-w)^2/(tw)}-100r_t-128 r_z+36 r_w+333+9 \pi ^2\right]
\nonumber\\
&-&
\frac{g^2 G^4}{96} \left[219 r_z^2-40r_{t/z}^2+21 r_{w/z}^2-730 r_z+6 r_w+715+200 \frac{\pi ^2}{3}\right]
\nonumber\\
&+&\frac{2}{3}G^2 y_t^4 \left(3r_t^2-8 r_t+9\right)-
\frac{G^6}{192} \left(34 r_{t/z}^2-273 r_z^2+3 r_{w/z}^2+940 r_z-961-206 \frac{\pi ^2}{3}\right)
\nonumber
\eea
\bea
&+&
\frac{G^4 y_t^2}{48} \left[27 \left(r_{t/z}^2-r_z^2\right)-68 r_t-28 r_z +189\right]
+\frac{5}{3}g^2G^2 y_t^2\left(2 r_t+4 r_z-9\right)
\nonumber\\
&-&\left.
\frac{3y_t^6 }{2}\left(3 r_t^2+2r_{t/w}r_{(t-w)/t} -16 r_t+23+\frac{\pi ^2}{3}
\right)+\frac{3}{4} \left(g^6-3 g^4 y_t^2+4 y_t^6\right)\mathrm{Li}_2[w/t]\right.
\nonumber\\
&+&
\frac{y_t^2}{48}\left[\left(14G^2- 160g^2+ 128\frac{g^4}{G^2}\right) y_t^2 +17G^4 -40g^2G^2+ 32g^4\right] \xi_{11zt}
\nonumber\\
&+&\left.
\frac{g^2}{192}\left[3 G^4+4\left(12G^2-51g^2-36\frac{g^4}{G^2}\right)g^2\right]\xi_{11zw}
\right\}\ ,
\label{lambdaeff2}
\eea
where $\xi_{11xy}=\xi(1,1,x/y)$,
\be
r_p\equiv\ln[\kappa_pe^{2\gamma(h)}]\ ,\;\;
r_{t/w}\equiv \ln[\kappa_t/\kappa_w]\ ,\;\; 
r_{(t-w)/t}\equiv \ln[(\kappa_t-\kappa_w)/\kappa_t]\ ,\;\; 
\ee
and so on. It can be checked that (\ref{lambdaeff2}) reproduces (\ref{lambdaeff})
in the (electroweak)  gaugeless limit $g,g'\rightarrow 0$.


\small



\begin{thebibliography}{99}

\bibitem{ATLAS:2012ac}
ATLAS Collaboration,
  Phys.\  Lett.\ B{710} (2012) 49
  [arXiv:1202.1408].
  F. Gianotti, the ATLAS Collaboration, talk given at CERN on July 4, 2012.


%

\bibitem{Chatrchyan:2012tx}
CMS Collaboration, Phys. Lett. B710 (2012) 26
  [arXiv:1202.1488].
 J. Incandela, the CMS Collaboration, talk given at CERN on July 4, 2012.



\bibitem{R1} N.~Cabibbo, L.~Maiani, G.~Parisi and R.~Petronzio,
  Nucl.\ Phys.\  B {158}(1979) 295. 
\bibitem{R2}  
  P.~Q.~Hung,
  Phys.\ Rev.\ Lett.\  {42} (1979)  873.
\bibitem{R3}    
  M.~Lindner,
  Z.\ Phys.\  C {31} (1986)  (1986).
  
\bibitem{R4}     
M.~Sher,
  Phys.\ Rept.\  {179} (1989) 273.
\bibitem{R5}   
B.~Schrempp and M.~Wimmer,
  Prog.\ Part.\ Nucl.\ Phys.\  {37} (1996) 1.  
\bibitem{R6} G.~Altarelli and G.~Isidori,
  Phys.\ Lett.\  B {337} (1994) 141.



\bibitem{CEQ}
  J.~A.~Casas, J.~R.~Espinosa and M.~Quir\'os,
  Phys.\ Lett.\  B {342} (1995) 171;
  Phys.\ Lett.\  B {382} (1996) 374.

 
\bibitem{IRS}
  G.~Isidori, G.~Ridolfi and A.~Strumia,
  Nucl.\ Phys.\ B\ {609} (2001) 387
  [hep-ph/0104016].
 
\bibitem{BdCE}
  C.~P.~Burgess, V.~Di Clemente and J.~R.~Espinosa,
  JHEP {0201} (2002) 041
  [hep-ph/0201160].
 

\bibitem{Isidori:2007vm}
   G.~Isidori, V.~S.~Rychkov, A.~Strumia and N.~Tetradis,
  Phys.\ Rev.\ D\ {77} (2008) 025034
  [hep-ph/0712.0242].
 
 \bibitem{ArkaniHamed:2008ym}  
     N.~Arkani-Hamed, S.~Dubovsky, L.~Senatore and G.~Villadoro,
  JHEP {0803} (2008) 075
  [arXiv:0801.2399].

\bibitem{Bezrukov:2009db}
  F.~Bezrukov and M.~Shaposhnikov,
  JHEP {0907} (2009) 089
  [hep-ph/0904.1537].
 
 

 
 \bibitem{Hall:2009nd}
  L.~J.~Hall and Y.~Nomura,
  JHEP {1003} (2010) 076
  [arXiv:0910.2235].
 
    
\bibitem{EEGHR}
    J.~Ellis, J.~R.~Espinosa, G.~F.~Giudice, A.~Hoecker and A.~Riotto,
  Phys.\ Lett.\ B\ {679} (2009) 369
  [hep-ph/0906.0954].


\bibitem{EliasMiro:2011aa}
  J.~Elias-Miro, J.~R.~Espinosa, G.~F.~Giudice, G.~Isidori, A.~Riotto and A.~Strumia,
  Phys.\ Lett.\ B {709} (2012) 222
  [hep-ph/1112.3022].
  

\bibitem{betagauge3}
L.~N.~Mihaila, J.~Salomon and M.~Steinhauser,
  Phys.\ Rev.\ Lett.\  {108} (2012) 151602
  [arXiv:1201.5868].

 
\bibitem{Chetyrkin:2012rz}
   K.~G.~Chetyrkin and M.~F.~Zoller,
arXiv:1205.2892.
  
\bibitem{Bezrukov:2012sa}
  F.~Bezrukov, M.~Y.~Kalmykov, B.~A.~Kniehl and M.~Shaposhnikov,
  [hep-ph/1205.2893].

 
\bibitem{Nielsen}
D.L.~Bennett, H.B.~Nielsen and I.~Picek, \plb{208}{1988}{275};
C.D.~Froggatt and H.B.~Nielsen, \plb{368}{1996}{96}.


\bibitem{Shaposhnikov:2009pv}
  M.~Shaposhnikov and C.~Wetterich,
  Phys.\ Lett.\ B {683} (2010) 196
  [hep-ph/0912.0208].

 
\bibitem{Holthausen:2011aa}
  M.~Holthausen, K.~S.~Lim and M.~Lindner,
  JHEP {1202} (2012) 037
  [arXiv:1112.2415].


 \bibitem{Masina:2011aa}
  I.~Masina and A.~Notari,
 arXiv:1112.2659.
 
 \bibitem{Masina:2012yd}
  I.~Masina and A.~Notari,
arXiv:1204.4155.

 
\bibitem{V2}
  C.~Ford, I.~Jack and D.~R.~T.~Jones,
  Nucl.\ Phys.\ B {387} (1992) 373
   [Erratum-ibid.\ B {504} (1997) 551]
  [hep-ph/0111190].
See also, 
  S.~P.~Martin,
  Phys.\ Rev.\ D {65} (2002) 116003
  [hep-ph/0111209].
  
\bibitem{Martinself}
  S.~P.~Martin,
  Phys.\ Rev.\ D {70} (2004) 016005
  [hep-ph/0312092].

\bibitem{Martinint}
  S.~P.~Martin,
  Phys.\ Rev.\ D {68} (2003) 075002
  [hep-ph/0307101].
  

  
\bibitem{SZ}
  A.~Sirlin and R.~Zucchini,
  Nucl.\ Phys.\ B {266} (1986) 389.

\bibitem{DS}
  G.~Degrassi and P.~Slavich,
  JHEP {1011} (2010) 044
  [arXiv:1007.3465].


\bibitem{Fleischer:1999hp}
  J.~Fleischer, M.~Y.~.Kalmykov and A.~V.~Kotikov,
  Phys.\ Lett.\ B {462} (1999) 169
  [hep-ph/9905249].





  \bibitem{alpha}
Particle Data Group,
  J.\ Phys.\ G {37} (2010) 075021.
  The LEP Electroweak Working Group, \url{http://lepewwg.web.cern.ch}.
  We thank Jens Erler and Paul Langacker for the latest
  fit we quote, and  Martin Gr\"unewald for useful discussions.
 
\bibitem{alpha3} 
  S.~Bethke,
  Eur.\ Phys.\ J.\ C {64} (2009) 689
  [hep-ph/0908.1135].
 
 
\bibitem{topmass}
  Tevatron Electroweak Working Group,
  [hep-ex/1107.5255].
  
  \bibitem{MtCMS}
CMS collaboration,  
\href{http://cdsweb.cern.ch/record/1427762}{CMS-PAS-TOP-11-015}.

\bibitem{MtATLAS}
ATLAS collaboration, \href{http://cdsweb.cern.ch/record/1441190/files/ATL-PHYS-SLIDE-2012-106.pdf}{Top quark mass measurements at the ATLAS experiment}.
  
  

\bibitem{Broadhurst:1991fy}
  D.~J.~Broadhurst, N.~Gray and K.~Schilcher,
  Z.\ Phys.\ C {52} (1991) 111.
      
\bibitem{Melnikov:2000qh}
  K.~Melnikov and T.~v.~Ritbergen,
  Phys.\ Lett.\ B {482} (2000) 99
  [hep-ph/9912391];
  K.~G.~Chetyrkin and M.~Steinhauser,
  Nucl.\ Phys.\ B {573} (2000) 617
  [hep-ph/9911434].

\bibitem{Hempfling:1994ar}
  R.~Hempfling and B.~A.~Kniehl,
  Phys.\ Rev.\ D {51} (1995) 1386
  [hep-ph/9408313].
  
  
\bibitem{Jegerlehner:2003py}
  F.~Jegerlehner and M.~Y.~.Kalmykov,
  Nucl.\ Phys.\ B {676} (2004) 365
  [hep-ph/0308216].
  
 
\bibitem{Hoang:2008xm}
  A.~H.~Hoang and I.~W.~Stewart,
  Nucl.\ Phys.\ Proc.\ Suppl.\  {185} (2008) 220
  [hep-ph/0808.0222].

\bibitem{R7}  
S. Fleming, A. H. Hoang, S. Mantry and I. W. Stewart, 
Phys.\ Rev.\ D 77 (2008) 074010 [hep-ph/0703207]. 

\bibitem{R7b}  
S.~Moch, P.~Uwer and A.~Vogt,  Phys.\ Lett.\ B {714} (2012) 48  [arXiv:1203.6282]. 

\bibitem{R8}  
  S.~Alekhin, A.~Djouadi and S.~Moch,
  Phys.\ Lett.\ B {\bf 716} (2012) 214
  [arXiv:1207.0980].



\bibitem{Veltman:1980mj}
  M.~J.~G.~Veltman,
  Acta Phys.\ Polon.\ B {12} (1981) 437.


\bibitem{Bezrukov:2007ep}
  F.~L.~Bezrukov and M.~Shaposhnikov,
  Phys.\ Lett.\ B {659} (2008) 703
  [hep-ph/0710.3755].

\bibitem{Burgess:2009ea}
  C.~P.~Burgess, H.~M.~Lee and M.~Trott,
  JHEP {0909} (2009) 103
  [hep-ph/0902.4465];
  J.~L.~F.~Barbon and J.~R.~Espinosa,
  Phys.\ Rev.\ D {79} (2009) 081302
  [hep-ph/0903.0355].
  
\bibitem{Lerner:2010mq}
  R.~N.~Lerner and J.~McDonald,
  Phys.\ Rev.\ D {82} (2010) 103525
  [hep-ph/1005.2978];
    G.~F.~Giudice and H.~M.~Lee,
  Phys.\ Lett.\ B {694} (2011) 294
  [hep-ph/1010.1417].
  
\bibitem{DeSimone:2008ei}
  A.~De Simone, M.~P.~Hertzberg and F.~Wilczek,
  Phys.\ Lett.\ B {678} (2009) 1
  [hep-ph/0812.4946].
    
\bibitem{EliasMiro:2012ay}
  J.~Elias-Miro, J.~R.~Espinosa, G.~F.~Giudice, H.~M.~Lee and A.~Strumia,
  [hep-ph/1203.0237].

\bibitem{Giudice:2011cg}
  G.~F.~Giudice and A.~Strumia,
  Nucl.\ Phys.\ B {858} (2012) 63
  [hep-ph/1108.6077].


\bibitem{CCD}
  M.~E.~Cabrera, J.~A.~Casas and A.~Delgado,
  Phys.\ Rev.\ Lett.\  {108} (2012) 021802
  [hep-ph/1108.3867].
  
  \bibitem{split}
  N.~Arkani-Hamed and S.~Dimopoulos,
  JHEP {0506} (2005) 073
  [hep-th/0405159];
    G.~F.~Giudice and A.~Romanino,
  Nucl.\ Phys.\ B {699} (2004) 65
  [hep-ph/0406088];
    N.~Arkani-Hamed, S.~Dimopoulos, G.~F.~Giudice and A.~Romanino,
  Nucl.\ Phys.\ B {709} (2005) 3
  [hep-ph/0409232].
  
  \bibitem{anmod}
    G.~F.~Giudice, M.~A.~Luty, H.~Murayama and R.~Rattazzi,
  JHEP {9812} (1998) 027
  [hep-ph/9810442];
    J.~D.~Wells,
  Phys.\ Rev.\ D {71} (2005) 015013
  [hep-ph/0411041];
    N.~Arkani-Hamed, A.~Delgado and G.~F.~Giudice,
  Nucl.\ Phys.\ B {741} (2006) 108
  [hep-ph/0601041].
  
  \bibitem{Giudice:2006sn}
  G.~F.~Giudice and R.~Rattazzi,
  Nucl.\ Phys.\ B {757} (2006) 19
  [hep-ph/0606105].
  
  \bibitem{SOC}
   P.~Bak, C.~Tang and K.~Wiesenfeld,  Phys.\ Rev.\ Lett.\  {59} (1987) 381.
 



      
\bibitem{CW}
  S.~R.~Coleman and E.~J.~Weinberg,
  Phys.\ Rev.\ D {7} (1973) 1888.
      
      
\end{thebibliography}
\end{document}